\documentclass[%
 reprint,superscriptaddress,
 amsmath,amssymb,prb,
]{revtex4-2}

\usepackage{graphicx}% Include figure files
\usepackage{dcolumn}% Align table columns on decimal point
\usepackage{bm}% bold math
\usepackage{color}
\usepackage{siunitx}
\usepackage[version=3]{mhchem} % Formula subscripts using \ce{}

\usepackage{xcolor}
\newcommand{\tred}[1]{\textcolor{black}{#1}}

\def \yc{\ce{YbBi2ClO4}}
\def \yi{\ce{YbBi2IO4}}
\def \yx{YbBi$_2$XO$_4$}
\def \degC{$^{\circ}$C~}
\def \TN{$T_{\mathrm N}$}
\def \CW{$\theta_{\mathrm {CW}}$}

\begin{document}

\preprint{APS/123-QED}

\title{Quantum magnetism in the frustrated square lattice oxyhalides \yi{} and \yc{}} \thanks{This manuscript has been authored by UT-Battelle, LLC under Contract No. DE-AC05-00OR22725 with the U.S. Department of Energy.  The United States Government retains and the publisher, by accepting the article for publication, acknowledges that the United States Government retains a non-exclusive, paid-up, irrevocable, world-wide license to publish or reproduce the published form of this manuscript, or allow others to do so, for United States Government purposes.  The Department of Energy will provide public access to these results of federally sponsored research in accordance with the DOE Public Access Plan (http://energy.gov/downloads/doe-public-access-plan).}

\author{Pyeongjae Park}
\email{parkp@ornl.gov}
\affiliation{Materials Science \& Technology Division, Oak Ridge National Laboratory, Oak Ridge, TN 37831, USA}

\author{G. Sala}
%\email{salag@ornl.gov}
\affiliation{Spallation Neutron Source, Second Target Station, Oak Ridge National Laboratory, Oak Ridge, TN, 37831, USA}

\author{Th. Proffen}
\affiliation{Neutron Scattering Division, Oak Ridge National Laboratory, Oak Ridge, Tennessee 37831, USA}

\author{Matthew B. Stone}
\affiliation{Neutron Scattering Division, Oak Ridge National Laboratory, Oak Ridge, Tennessee 37831, USA}

\author{Andrew D. Christianson}
\affiliation{Materials Science \& Technology Division, Oak Ridge National Laboratory, Oak Ridge, TN 37831, USA}

\author{Andrew F. May}
\email{mayaf@ornl.gov}
\affiliation{Materials Science \& Technology Division, Oak Ridge National Laboratory, Oak Ridge, TN 37831, USA}

%\date{\today}% It is always \today, today, but any date may be explicitly specified

\begin{abstract}

Square-lattice systems offer a direct route for realizing 2D quantum magnetism with frustration induced by competing interactions.  In this work,  the square-lattice materials \yi{} and \yc{} were investigated using a combination of magnetization and specific heat measurements on polycrystalline samples.  Specific heat measurements provide evidence for long-range magnetic order below \TN{} = 0.21\,K (0.25\,K) for \yi{} (\yc{}).  On the other hand, a rather broad maximum is found in the temperature-dependent magnetic susceptibility, located at $T_{\mathrm{max}}$ = 0.33\,K (0.38\,K) in \yi{} (\yc{}), consistent with the quasi-2D magnetism expected for the large separation between the magnetic layers.  Estimation of the magnetic entropy supports the expected  Kramers’ doublet ground state for Yb$^{3+}$ and the observed paramagnetic behavior is consistent with a well-isolated doublet.  Roughly two-thirds of the entropy is consumed above \TN{}, due to a combination of the quasi-2D behavior and magnetic frustration.  The impact of frustration is examined from the viewpoint of a simplified $J_1$-$J_2$ square lattice model, which is frustrated for antiferromagnetic interactions.  Specifically, a high-temperature series expansion analysis of the temperature-dependent specific heat and magnetization data yields $J_2$/$J_1$ = 0.30 (0.23) for \yi{} (\yc{}).  This simplified analysis suggests strong frustration that should promote significant quantum fluctuations in these compounds, and thus motivates future work on the static and dynamic magnetic properties of these materials.

\end{abstract}

\maketitle

\section{Introduction}
The $S = 1/2$ square lattice antiferromagnet has served as a pivotal spin model for understanding quantum magnetism in two dimensions (2D) \cite{Reger_1988, Dagotto_1989, QMC_1991, Schulz_1992,Chandra_1988}. Intense interest in this model emerged from its relevance to high-temperature superconducting cuprates, in which the Cu$^{2+}$ ions form a quasi-2D square lattice structure of spin-1/2 moments \cite{RMP_1991, Reger_1988}. It was proposed that the inherent quantum spin fluctuations within this spin system contribute to the formation of superconductivity in cuprates. This was naturally extended to the suggestion of a new quantum-entangled magnetic ground state devoid of classical long-range order \cite{Anderson_1987}, now called a quantum spin liquid (QSL). Although it is now established that a $S = 1/2$ square lattice with nearest-neighbor (NN) antiferromagnetic interactions ($J_{1}$) develops long-range order, quantum fluctuations still play a role in this system \cite{Reger_1988, Hamer_1992, Mourigal_2015, Christensen_2007}. This is notably evident through a large reduction in its ordered moment, which diminishes to approximately 60\% of the classical limit \cite{Reger_1988,Hamer_1992}. Additionally, spectroscopic studies of the spin dynamics of $S = 1/2$ square lattice materials have revealed indications of fractionalized quasi-particles \cite{Mourigal_2015} and quantum entanglement \cite{Christensen_2007}, making model systems in this class potential testing grounds for such behavior. 

 The importance of quantum effects in $S = 1/2$ square lattice antiferromagnets can be further amplified by adding frustration. This can be realized by antiferromagnetic second NN interactions ($J_{2}$).  In this case, competition between $J_{1}$ and $J_{2}$ introduces exchange frustration on a square lattice, which, combined with low spin and low dimension, substantially increases the influence of quantum fluctuations \cite{Reger_1988, Dagotto_1989, QMC_1991, Schulz_1992, Chandra_1988}. In accordance with this idea, theoretical works have suggested possible emergence of a QSL within a parameter range of $0.4 < J_{2}/J_{1} <$ 0.6 in which the frustration is maximized \cite{Balents_2012, Morita_2015, Gong_2014,Liu_2022_Tens}. Notably, this phase emerges between the regions of Neel-type ($J_{2}/J_{1} < 0.4$) and columnar-type ($0.6 < J_{2}/J_{1}$) antiferromagnetic orders, effectively bridging these two phases seamlessly by suppressing each order. Thus, the $S = 1/2$ $J_{1}$--$J_{2}$ square lattice model provides a simple but informative platform for studying 2D quantum magnetism where the degree of quantum fluctuations is systematically controlled by $J_{2}$. Consequently, continuous efforts have been made to find such frustrated $S = 1/2$ square lattice materials with sizable antiferromagnetic $J_{2}$ \cite{Sq_ex_1,Ref_Tmax2,Sq_ex_3,Sq_ex_4,Sq_ex_5,Ref_Tmax2}.

Halide based materials have risen as important sources for studying magnetic behavior in 2D lattices \cite{TM_halide_2017}. Such investigations range from studying genuine 2D magnetism in atomically-thin samples fabricated by mechanical exfoliation \cite{CrI3_nat,NiI2_nat,NiI2_bilayer,burch2018magnetism,huang2020emergent} to nearly ideal realizations of quantum magnetism \cite{RuCl3_sci, RuCl3_nphys, YbCl3_natc, CoI2_nphys}. Notably, the latter realizations were all achieved by spin-orbit coupled electron configurations (\textit{i.e.}, $J$-manifold) in which the Kramers' doublet ground states manifest an effective spin-1/2 ($J_{\mathrm{eff}}=1/2$) degree of freedom. This pseudospin mechanism has provided an important method of designing quantum magnets beyond cuprate (Cu$^{2+}$) or vanadate (V$^{4+}$) systems that have been the subjects of intense investigation \cite{qmag_review}. Particular success has been found in hexagonal systems such as honeycomb \ce{RuCl3} \cite{RuCl3_sci,RuCl3_nphys} and \ce{YbCl3} \cite{YbCl3_natc}, and triangular \ce{CoI2} \cite{CoI2_nphys}. However, $J_{\mathrm{eff}}=1/2$ sqaure lattice materials built upon halide chemistry are still lacking. 
%%%%%%%%%%%%%%%%%%%%%%% Figure - structure %%%%%%%%%%%%%%%%%%%%%%%%%%%%%%%%
\begin{figure}
\includegraphics[width=0.96\columnwidth]{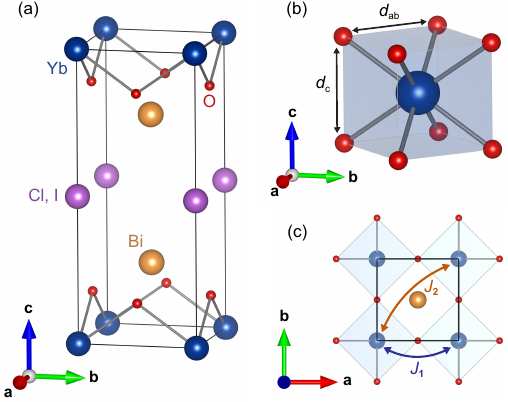} 
\caption{\label{crystal} Crystal structure of \yx{} (X=Cl, I). (a) The tetragonal unit cell of \yx{}, having the space group $P4/mmm$ (123). The Yb ions form a stack of square lattices separated by Bi and Cl (I) layers. (b) The oxygen coordination of the Yb ions. The distances between Yb and O are all equal. The ratio of in-plane to out-of-plane O--O distances ($d_{\mathrm{ab}}/d_{\mathrm{c}}$) are 0.996 for both \yi{} and \yc{}. (c) Nearest-neighbor ($J_{\mathrm{1}}$) and second nearest-neighbor ($J_{\mathrm{2}}$) exchange interactions of a Yb$^{3+}$ square lattice. Crystal structures were drawn with Vesta \cite{vesta}. }
\end{figure}
%%%%%%%%%%%%%%%%%%%%%%%%%%%%%%%%%%%%%%%%%%%%%%%%%%%%%%%%%%%%%%%%%%%%%%%%%%%%

In this article, we report the physical properties of the tetragonal materials \yi{} and \yc{} (Fig. \ref{crystal}), which are members of large class of oxy-halide magnetic materials \ce{ReBi2XO4} (Re = rare-earth elements, X = Cl, Br, I) \cite{Schmidt_2000_structure}. The Yb sites in these compounds form square lattices, with the separation between each layer more than twice as long as the nearest Yb--Yb distance.   
 The temperature-dependent magnetization precisely follows the Curie-Weiss behavior from 50\,K to $\sim$1\,K, with a negative Curie-Weiss temperature of \CW{} = -0.78\,K (-0.73\,K) in \yi{} (\yc{}). The specific heat ($C_{p}$) measurements reveal the ordering of Yb$^{3+}$ moments below \TN{} = 0.21\,K (0.25\,K) in \yi{} (\yc{}). The estimated total magnetic entropy ($\sim$$R\mathrm{ln}(2)$) indicates Yb$^{3+}$ in \yi{} and \yc{} manifests well-isolated Kramers' doublet ground states, consistent with the Curie-Weiss law being valid up to 50\,K\,$\sim$\,220\,\TN{}. Meanwhile, temperature-dependent magnetic susceptibilities exhibit rather broad maxima at temperatures ($T_{\mathrm{max}}$) higher than \TN{}. This evidences strong spin correlations above \TN{}, consistent with the observation that the majority of the magnetic entropy is released above \TN{} in these materials. Notably, both the $|$\CW$|$/$T_{\mathrm{max}}$ ratio and $J_{2}/J_{1}$ estimated from a high-temperature series expansion analysis suggest these compounds possess sizable frustration, indicating a possibility of strong quantum effects.

\section{Experimental details}

Polycrystalline \ce{YbBi2IO4} and \ce{YbBi2ClO4} were formed by reactions of \ce{Bi2O3},  \ce{Yb2O3}, and \ce{BiIO} or \ce{BiClO} in evacuated \ce{SiO2} tubing. \ce{YbBi2IO4} was obtained from a stoichiometric reaction at 950\degC for a total of 6\,d with one intermediate grinding.  For \ce{YbBi2ClO4}, after an initial reaction at 950\degC for 100\,h, the material was ground and heated at 900\degC for 100\,h. After this, impurities were detected by laboratory x-ray diffraction, and thus an additional $\approx$1\% of the original mass of \ce{BiClO} and \ce{Bi2O3} were added and the sample was heated at 850\degC for 150\,h, which resulted in the product used in the experiments reported herein. \ce{Yb2O3} was dried prior to use.  The source powders of \ce{BiIO} and \ce{BiClO} were formed by reacting stoichiometric mixtures of \ce{Bi2O3} and ultra-dry \ce{BiI3} spheres or anyhydrous \ce{BiCl3} at 550\degC for 24\,h with an intermediate dwell of 10\,h at 300-350\degC; the reagents were ground together in a glovebox and subsequently sealed under vacuum in \ce{SiO2} without exposure to air.  

Sample quality was initially checked using a commercial powder x-ray diffractometer (PANalytical X’Pert Pro MPD) with Cu--K$_{\alpha1}$ radiation ($\lambda$ = 1.5406\,\AA) from an incident beam monochromator. A full structural analysis was performed using neutron diffraction data collected at the POWGEN time-of-flight diffractometer at the Spallation Neutron Source (SNS) \cite{powgen2011}. For the measurements, a \ce{YbBi2IO4} (\ce{YbBi2ClO4}) powder sample with a mass of 6.09 g (3.73 g) was loaded in a standard vanadium can. The data were collected at 30 K with the high resolution mode and a center wavelength of 1.5 \AA. Rietveld refinement was performed using the FullProf software package \cite{fullprof}.

%%%%%%%%%%%%%%%%%%%%%%% Table -  structure parameters %%%%%%%%%%%%%%%%%%%%%%%%
{\renewcommand{\arraystretch}{1.25}
\begin{table*}[t]
\caption{\label{tab_refine} Crystal structure of YbBi$_2$(I,Cl)O$_4$ determined by Rietveld refinement of the powder neutron diffraction data collected at 30 K (Fig. \ref{NPD}). The space group is $P4/mmm$ (No. 123). The occupancies of all sites are equal to unity.}
\begin{ruledtabular}
\begin{tabular}{cccccccccccc}
 & & &\multicolumn{4}{c}{YbBi$_2$IO$_4$}& &\multicolumn{4}{c}{YbBi$_2$ClO$_4$}\\
 & & &\multicolumn{4}{c}{\textit{a} = \textit{b} = 3.8658(1) \si{\angstrom}, \textit{c} = 9.5052(1) \si{\angstrom}}& 
 &\multicolumn{4}{c}{\textit{a} = \textit{b} = 3.8196(1) \si{\angstrom}, \textit{c} = 8.8169(1) \si{\angstrom}}\\ \hline
 Atom&Site& &\textit{x/a}&\textit{y/b}&\textit{z/c}&\textit{B}$_{\mathrm{iso}}$ (\si{\angstrom}$^2$) & &\textit{x/a}&\textit{y/b}&\textit{z/c}&\textit{B}$_{\mathrm{iso}}$ (\si{\angstrom}$^2$) \\ \hline
 Yb&$1a$ &&0&0&0&0.057(14)& &0&0&0&0.041(10)\\
 (I,Cl)&$1b$& &0&0&0.5&0.148(27)& &0&0&0.5&0.235(13)\\
 Bi&$2h$& &0.5&0.5&0.7425(1)&0.038(10)& &0.5&0.5&0.7195(1)&0.002(11)\\
 O&$4i$& &0.5&0&0.8581(1)&0.236(10)& &0.5&0&0.8463(1)&0.193(12)\\ \hline
& & &\multicolumn{4}{c}{$R_{p}$(\%) = 8.26, $R_{wp}$(\%) = 6.86}& &\multicolumn{4}{c}{$R_{p}$(\%) = 5.23, $R_{wp}$(\%) = 5.85}\\
& & &\multicolumn{4}{c}{$R_{\mathrm{exp}}$(\%) = 0.722, $\chi^{2}$ = 90.4}& &\multicolumn{4}{c}{$R_{\mathrm{exp}}$(\%) = 0.793, $\chi^{2}$ = 54.4}\\
\end{tabular}
\end{ruledtabular}
\end{table*}
%%%%%%%%%%%%%%%%%%%%%%%%%%%%%%%%%%%%%%%%%%%%%%%%%%%%%%%%%%%%%%%%%%%%%%%%%%%%%

Magnetization and specific heat measurements were performed in Quantum Design measurement systems. Magnetization measurements were performed in an MPMS3 using a gelcap to hold the polycrystalline materials within a standard measurement straw. Data below 1.8\,K were obtained using the iHe-3 insert, for which samples were affixed to the straw by mixing them with high-vacuum grease. Specific heat measurements below 2.5\,K were conducted using a dilution refrigerator insert within a Dynacool. In order to improve the mechanical integrity and thermal equilibration of the samples, the powders were gently ground with an equal mass of high purity silver powder and then pressed into a small pellet. The contribution of the silver powder was subtracted using in-house data for silver's specific heat, which is much smaller than the magnetic contribution of the samples at low-temperatures. AC susceptibility measurements down to 100\,mK were performed using a dilution refrigerator insert with the ac-dr option from Quantum Design. A portion of the pellets used for specific heat (including Ag powder) was affixed to the sample holder using N-grease. Measurements were completed upon warming after thermalizing below 100 mK for several hours. Data were collected using 227 and 756 Hz and the broad maxima of susceptibility observed occurred at the same temperature for both frequencies; a static DC field was not applied and the ac driving amplitude was 2 Oe.

\section{Results}

\subsection{Structure analysis}

The crystal structures of \yi{} and \yc{} were studied through their time-of-flight neutron diffraction profiles collected at $T$ = 30\,K. Fig. \ref{NPD} show the data and Rietveld refinement results as a function of momentum transfer $|$\textbf{Q}$|$. Nuclear Bragg peaks of likely impurity phases (\ce{Yb2O3}, \ce{YbI2}, \ce{BiI3}, \ce{Bi2O3}, \ce{BiIO}, \tred{\ce{Yb(IO3)3}, \ce{YbBi2}, and \ce{Yb5Bi3}}) were not observed in the neutron diffraction profile. \tred{Yet we note that we observed two unidentified impurity peaks in the \yi{} data, located at $d$ = 1.84 and 2.13 \si{\angstrom}, with intensities less than 0.1 \% compared to those originating from \yi{}.} The refinements based on the previously-reported tetragonal unit cell ($P4/mmm$) \cite{Schmidt_2000_structure} yield an excellent agreement with the data, the results of which are summarized in Table \ref{tab_refine}. Notably, the chloride exhibits smaller thermal displacement parameters ($B_{\mathrm{iso}}$) of Yb, Bi, and O, which is especially noticeable on the Bi site. 

%%%%%%%%%%%%%%%%%%%%%%% Figure - neutron diffraction %%%%%%%%%%%%%%%%%%%%%%%%
\begin{figure}[h]
\includegraphics[width=1\columnwidth]{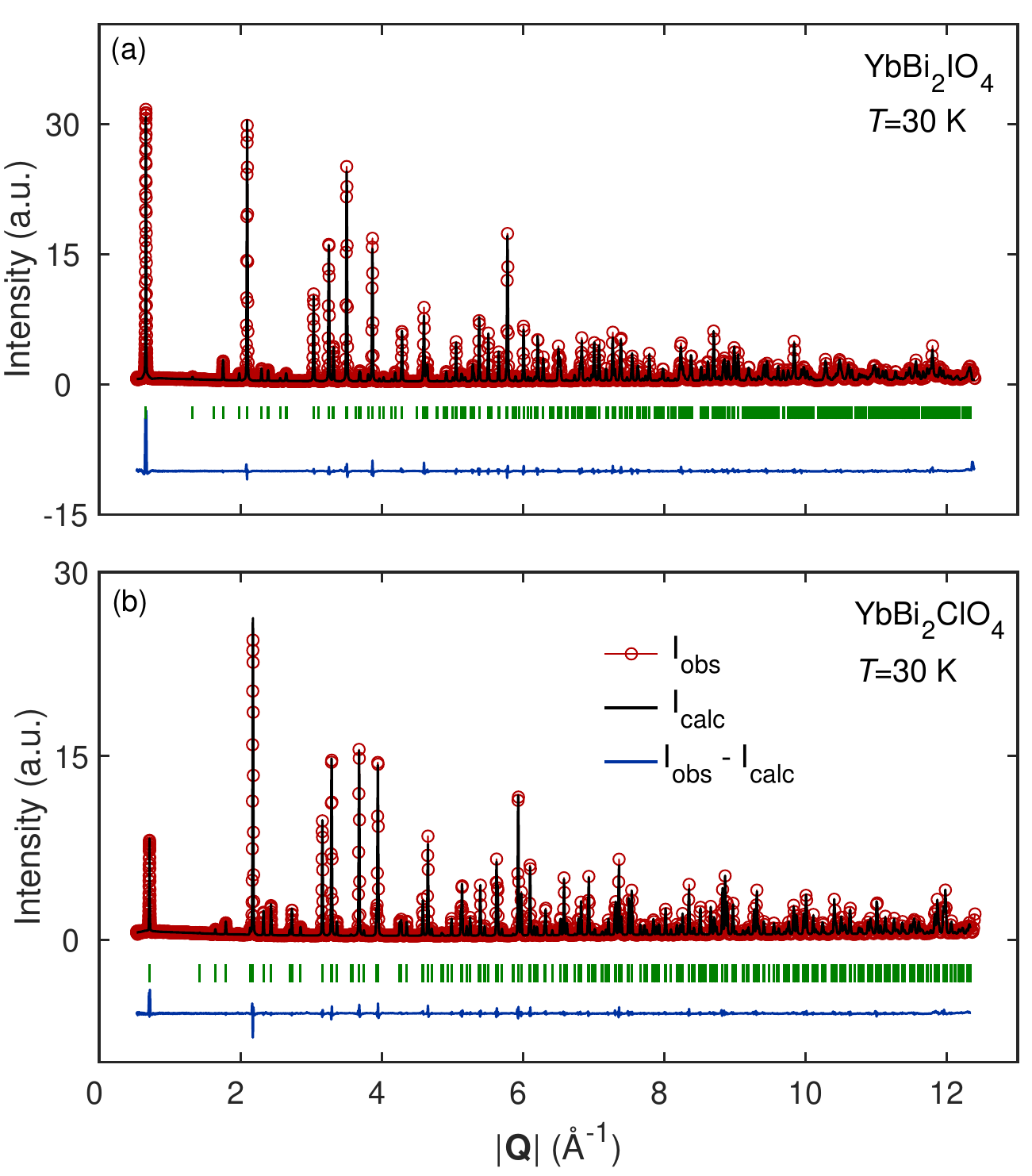} 
\caption{\label{NPD} Powder neutron diffraction profiles of (a) YbBi$_{2}$IO$_{4}$ and (b) YbBi$_{2}$ClO$_{4}$ collected at 30 K. Solid black lines in (a) and (b) are the simulated profile from the refinement result shown in Table \ref{tab_refine}. Solid blue lines are the difference between the data and simulations. Green vertical ticks denote the positions of nuclear reflections.}
\end{figure}
%%%%%%%%%%%%%%%%%%%%%%%%%%%%%%%%%%%%%%%%%%%%%%%%%%%%%%%%%%%%%%%%%%%%%%%%%%%%%

The magnetic Yb$^{3+}$ ions form a square lattice through edge sharing of square prismatic polyhedra as shown in Fig. \ref{crystal}(b)--(c). The NN distances are equal to the $a$ lattice parameter, and the second NN is on the diagonal at a distance of $\sqrt{2}a$. Each layer of the Yb$^{3+}$ square lattice is well isolated from each other, with a separation equal to the $c$ lattice parameter. As shown in Table \ref{tab_refine}, the ratio of in-plane to out-of-plane distances is greater than 2 ($c/a$ = 2.308 in \yc{} and $c/a$ = 2.459 in \yi), which drives the expectation for genuine quasi-2D magnetic behavior dominated by square lattice physics. This remains true even when comparing the out-of-plane distance with the second NN distance. The ratio ($c/\sqrt{2}a$ = 1.632 in \yc{} and 1.739 in \yi) is still comparable or even larger than the ratio of in-plane to out-of-plane bond distances in other quasi-2D square lattice antiferromagnets \cite{Mustonen_2018,Ref_Tmax2,Sq_ex_4,ZVPO}. Notably, the lattice parameter $c$ of \yi{} is nearly 8\% larger than that of \yc{}, while in-plane lattice parameters, differ by $\sim$1.2\% with the chloride having the smaller unit cell.

As expected from the space group, the oxygen coordination around Yb$^{3+}$ is highly symmetric (Fig. \ref{crystal}(b)). The eight O$^{2-}$ ligands are evenly spaced from Yb$^{3+}$ and form a nearly cubic-shaped polyhedron. Notably, the refined crystal structure of both \yi{} and \yc{} possesses a slight ($\sim$0.4\%) elongation of the polyhedron along the $c$-axis, as evident from the ratio of in-plane to out-of-plane O--O distances ($d_{\mathrm{ab}}/d_{\mathrm{c}}$) being smaller than 1 (Fig. \ref{crystal}(b)).

% \subsection{Crystal field level splitting}

\subsection{Magnetization and heat capacity measurements}

Fig. \ref{Fig4_pmag}(a)--(b) shows the temperature-dependent magnetization of \yi{} and \yc{} from 50 K to 0.4 K. Remarkably, \yi{} and \yc{} exhibit almost the same magnetization curve. No sign of a phase transition is found in this temperature range. As evident from the inset of Fig. \ref{Fig4_pmag}(a), the measured magnetization precisely follows the Curie-Weiss law for $2<T<50$\,K. This implies that the excited crystal field levels of Yb$^{3+}$ have a negligible contribution to the magnetism below 50 K. Orange solid lines in Fig. \ref{Fig4_pmag}(b) are the fit results with modified Curie-Weiss law ($\chi(T) = \chi_{0}+{C}/(T-\theta_{\mathrm{CW}})$, where $\theta_{\mathrm{CW}}$ denotes the Curie-Weiss temperature) for a temperature range from 2\,K to 20\,K. The fitting yielded \CW$=-0.78(3)$\,K and $\mu_{\mathrm{eff}} = 3.15(1)\mu_{\mathrm{B}}$ for \yi{}, and  \CW$=-0.73(3)$\,K and $\mu_{\mathrm{eff}} = 3.12(1)\mu_{\mathrm{B}}$ for \yc{}. Note that the fitted $\chi_{0}$ was less than $\mathrm{10^{-3}\,emu\:mol^{-1}\:Oe^{-1}}$ for both compounds. As expected from the magnitude of fitted \CW{}, the measured $M/H$ deviates from the Curie-Weiss behavior below around 0.8 K; see the inset of Fig. \ref{Fig4_pmag}(b).

The isothermal magnetization data shown in Fig. \ref{Fig4_pmag}(c) provide further information of the Yb$^{3+}$ magnetism in \yi{} and \yc{}. At 0.4\,K, full polarization of magnetic moments is achieved by applying a magnetic field up to 70 kOe. The saturated magnetization estimated from the high-field data yields $\sim$1.8$\mu_{\mathrm{B}}/\mathrm{Yb}^{3+}$ for each compound. For $J_{\mathrm{eff}}=1/2$ (as will become clear in the following paragraphs), this indicates the averaged $g$-factor of $\sim$3.6 for each compound, which is consistent with the magnitude of $\mu_{\mathrm{eff}}= g\sqrt{J(J+1)}\,\,(J=1/2)$ obtained from the Curie-Weiss fitting assumed to occur within the doublet.

%%%%%%%%%%%%%%%%%%%%%%% Figure - paramagnet %%%%%%%%%%%%%%%%%%%%%%%%
\begin{figure}
\includegraphics[width=0.9\columnwidth]{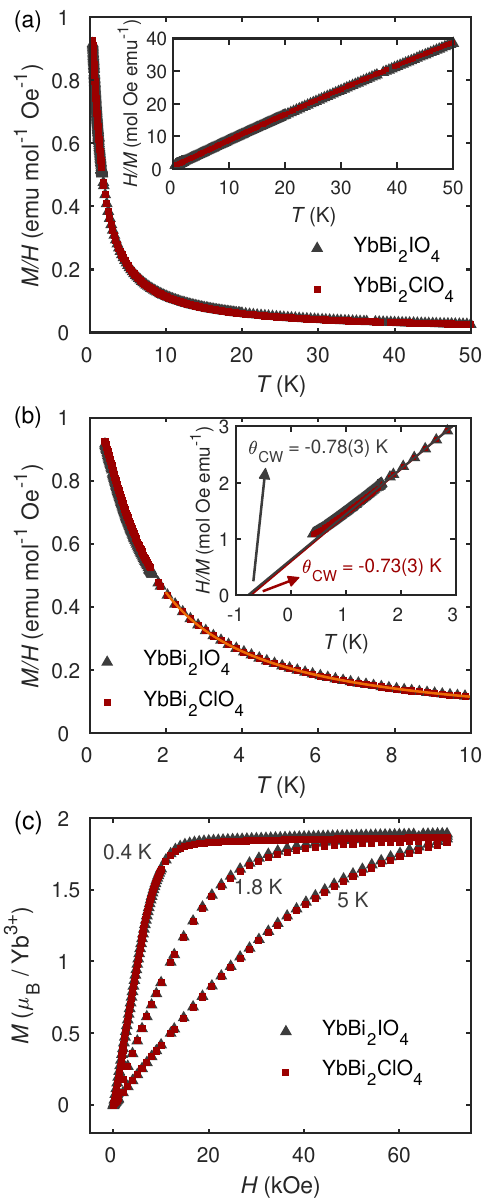} 
\caption{\label{Fig4_pmag} Magnetic properties of \yi{} and \yc{} above 0.4\,K ($T > $ \TN{}). (a)--(b) Temperature-dependent magnetization data measured down to 0.4\,K under a 10\,kOe external magnetic field. The insets show the inverse magnetic susceptibility, demonstrating (a) the Curie-Weiss behavior below 50 K and (b) deviation from the Curie-Weiss behavior below around 0.8\,K.  Solids lines in (b) are fit to the modified Curie-Weiss law between 2\,K and 20\,K. (b) Isothermal magnetization data from combined $^{3}$He and $^{4}$He experiments.}
\end{figure}
%%%%%%%%%%%%%%%%%%%%%%%%%%%%%%%%%%%%%%%%%%%%%%%%%%%%%%%%%%%%%%%%%%%%%%%%%%%%%

%%%%%%%%%%%%%%%%%%%%%%% Figure - dilution %%%%%%%%%%%%%%%%%%%%%%%%
\begin{figure}
\includegraphics[width=0.9\columnwidth]{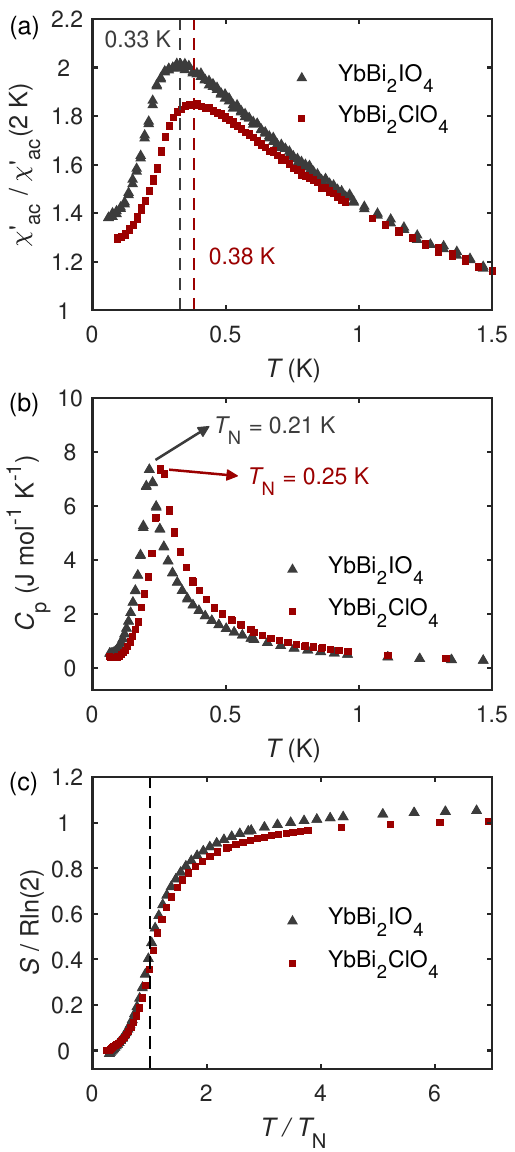} 
\caption{\label{Fig5_lowT} Phase transitions and magnetic correlations found for $T<0.4$\,K.  (a) Temperature-dependent ac susceptibility ($f$ = 756\,Hz) measured down to 100 mK. (b) Heat capacity ($C_{\mathrm P}$) of \yi{} and \yc{} around their Neel temperatures \TN{} = 0.21\,K and 0.25\,K, respectively. (c) The entropy obtained by direct integration of measured $C_{\mathrm P}/T$ without any background subtraction.}
\end{figure}
%%%%%%%%%%%%%%%%%%%%%%%%%%%%%%%%%%%%%%%%%%%%%%%%%%%%%%%%%%%%%%%%%%%%%%%%%%%%%

Upon further cooling, we observed clear signs of magnetic phase transitions in \yi{} and \yc{}. First, a broad peak appears in the real part of the ac magnetic susceptibility ($\chi^{'}_{\mathrm{ac}}$) measured down to $\sim$0.1\,K in a dilution refrigerator (Fig. \ref{Fig5_lowT}(a)). This peak is centered at $T_{\mathrm{max}}=0.33$\,K for \yi{} and $T_{\mathrm{max}}=0.38$\,K for \yc{}. This behavior is associated with short-range correlations above \TN{} due to the 2D character of the magnetism. In addition, no meaningful signal was found in the imaginary part ($\chi^{''}_{\mathrm{ac}}$), implying a fully compensated antiferromagnetic spin structure \tred{(\textit{i.e.}, zero net magnetization)} of \yi{} and \yc{} \cite{Ref_AC}. 

On the other hand, the heat capacity measurements are characterized by a sharp $\lambda$-shaped peak at 0.21\,K (0.25\,K) for \yi{} (\yc), which suggests the onset of long-range order in these compounds (Fig.\ref{Fig5_lowT}(b)). These transition temperatures (\TN) are $\sim$1.54 times lower than $T_{\mathrm{max}}$ found in the ac susceptiblity. Notably, the \TN{}  of \yc{} is 19 \% higher than that of \yi, consistent with the observation that $T_{\mathrm{max}}$ was 15 \% higher in \yc{}.

The magnetic entropy across the phase transition was estimated by integrating $C_{p}/T$ and the results are shown in Fig. \ref{Fig5_lowT}(c). Due to the very low temperatures involved, the phonon contribution to the entropy is assumed to be negligible in Fig. \ref{Fig5_lowT}(c). The entropy saturates to $\approx$$R\mathrm{ln}(2)\, (R=8.314\,\,\mathrm{J\,K^{-1}mol^{-1}})$, suggesting a well-isolated Kramers' doublet ground state that gives rise to $J_{\mathrm{eff}}=1/2$ magnetic moments. This isolation is also evident in the fact that both \yi{} and \yc{} follow the Curie-Weiss law up to 50\,K ($\sim$\,220\,\TN{}); thermal populations of higher-energy crystal field levels should be nearly zero around a few kelvin. In addition to supporting the $J_{\mathrm{eff}}=1/2$ picture in these compounds, the entropy calculation reveals that nearly two-thirds of the total magnetic entropy is consumed above \TN{}. This indicates the formation of sizable spin-spin correlation at a temperature 3--4 times higher than \TN{}, consistent with the deviation from the Curie-Weiss behavior below 0.8\,K (Fig. \ref{Fig4_pmag}(a)). 

\section{Discussion}

The size of \CW{}, $T_{\mathrm{max}}$, and \TN{} in \yi{} and \yc{} merits a deeper quantitative comparison as a diagnostic of whether these systems are frustrated. This is because $T_{\mathrm{max}}$ and \TN{} are suppressed by frustration whereas \CW{} simply captures a sum of all exchange interactions in a spin system (\textit{i.e.}, \CW{}$\,\propto(J_{1}+J_{2})$ for the $J_{1}$--$J_{2}$ model). First, even though the ratio $|$\CW$|$/\TN $= 3.71\:(2.92)$ found in \yi{} (\yc{}) largely surpasses 1, this alone should not be deemed as an indication of frustration in a 2D square lattice antiferromagnet. This is because \TN{} also depends on the size of inter-layer coupling ($J_{c}$) and thus $|$\CW$|$/\TN{} larger than 1 can occur without any frustration. Previous experimental studies of quasi-2D square lattice antiferromagnets demonstrate such a case: while they reported $|$\CW$|$/\TN{} $\sim$ 2.75 for \ce{Sr2CuTeO6} and $|$\CW$|$/\TN{} $\sim$ 1.7 for \ce{Zn2VO(PO4)2}, both compounds are known to have very small $J_{2}/J_{1}$ (\textit{i.e.,} very little frustration) \cite{Sq_ex_3,ZVPO}.

Instead, $|$\CW$|$/$T_{\mathrm{max}}$ better estimates the degree of frustration as $T_{\mathrm{max}}$ stems from the spin correlation of a 2D square lattice. Indeed, this ratio was reported to be close to 1 for both \ce{Sr2CuTeO6} and \ce{Zn2VO(PO4)2} \cite{Sq_ex_3,ZVPO}, and quantum Monte Carlo simulations yield $|$\CW$|$/$T_{\mathrm{max}}$ = 1.069 (close to 1) for $J_{2}=0$ in the $J_{1}$--$J_{2}$ model \cite{Ref_Tmax1,Ref_Tmax2}. Interestingly, the $|$\CW$|$/$T_{\mathrm{max}}$ values of \yi{} and \yc{} are 2.36 and 1.92, respectively, suggesting both \yi{} and \yc{} possess frustration. This frustration, in addition to the prounounced 2D character, may have further contributed to the significant spin-spin correlations above \TN{} in YbBi$_{2}$(I,Cl)O$_{4}$. However, it is important to note that a fitted \CW{} value can be influenced by various extrinsic factors in the measurement, necessitating more concrete evidence of frustration in \yi{} and \yc{}.

%%%%%%%%%%%%%%%%%%%%%%% Figure - highT fit %%%%%%%%%%%%%%%%%%%%%%%%
\begin{figure}
\includegraphics[width=1.0\columnwidth]{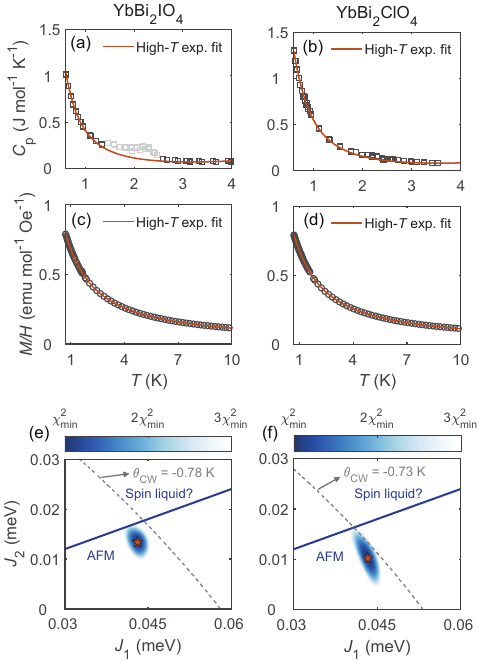} 
\caption{\label{Fig6_hFit} Quantitative analysis of the first and second NN exchange interactions using a high-temperature series expansion methodology. (a)--(d) Best fit of the measured $C_{p}$ and $\chi(T)$ with the high-temperature series expansion. The grayed data points in (a) contain a putative impurity signal and thus are excluded from the fitting. \tred{A short note on this impurity is provided in the main text.} (e)--(f) Two-dimensional color plots of the goodness-of-fit ($\chi^{2}$) as a function of $J_{1}$ and $J_{2}$. Orange star symbols denote $J_{1}$ and $J_{2}$ of the best fits ($\chi^{2}=\chi_{\mathrm{min}}^{2}$) shown in (a)--(d). Blue solid lines in (e)--(f) are the boundary between the Neel-type antiferromagnetic order and the spin liquid phase ($J_{2}=0.4J_{1}$). Grey dashed lines in (e)--(f) indicate the ($J_{1}$, $J_{2}$) parameter sets that yield \CW{}$=-0.78$\,K and -0.73\,K in $\chi(T)$ of the high-temperature series expansion.}
\end{figure}
%%%%%%%%%%%%%%%%%%%%%%%%%%%%%%%%%%%%%%%%%%%%%%%%%%%%%%%%%%%%%%%%%%%%%%%%%%%%%

To further explore the possibility of frustrated interactions, we conducted a high-temperature series expansion analysis based on the $S=1/2$ $J_{1}$--$J_{2}$ Heisenberg model frequently used for frustrated square lattices. This was done by performing a dual fit of measured $C_{p}(T)$ and $\chi(T)$ with their theoretical form derived by the high-temperature series expansion. We assumed $J_{1}$ is larger than $J_{2}$ as it is more likely from a bond length perspective. The resultant $J_{1}$ and $J_{2}$ values enable us to estimate the degree of frustration for each compound based on $J_{2}/J_{1}$. We used the analytic forms derived by Ref. \cite{highT_exp} that include polynomials up to the eighth order. The temperature range for the fitting was set to $0.6$\,K $< T < 4$\,K for $C_{p}$; the magnetic contribution above 4 K is nearly zero and the lattice contribution is negligible below 4 K. For $\chi(T)$, we used $0.6$\,K $ < T < 10$\,K. Note that the high-temperature series expansion is valid only for a temperature range higher than the thermodynamic energy scale of a spin system, \textit{i.e.}, $k_{\mathrm{B}}T >$ $S(S+1)\sqrt{{J_{1}}^{2} + {J_{2}}^{2}}$ \cite{highT_exp}. As will become clear in the following paragraph, 0.6 K is much higher than the thermodynamic energy scale of \yi{} and \yc{} obtained from this analysis, thereby satisfying the aforementioned condition. Calculated $C_{p}(T)$ and $\chi(T)$ were multiplied by scale factors to match their magnitude with our data (Fig. \ref{Fig6_hFit}(a)--(d)). In other words, our analysis only fitted the temperature dependence of $C_{p}(T)$ and $\chi(T)$, which is sensitive enough to discern the optimal values of $J_{1}$ and $J_{2}$. 

\tred{Before presenting the results of our series expansion analysis, we would like to briefly mention a potential impurity signal in $C_{p}(T)$ of \yi{} around 2.2 K (Fig. \ref{Fig6_hFit}(a)). The most likely impurity candidate is Yb$_{2}$O$_{3}$ as it undergoes magnetic phase transition at 2.25 K \cite{Yb2O3}. However, as mentioned earlier in this article, our powder neutron diffraction data (Fig. \ref{NPD}) revealed no nuclear Bragg peaks of Yb$_{2}$O$_{3}$. This observation raises two possibilities: i) the quantity of \ce{Yb2O3}, though not negligible, might be exceedingly small, or ii) it falls outside the range of considered impurity candidates, possibly related to the two unidentified impurity peaks found in our neutron diffraction data (see Section III. A.)}

Fig. \ref{Fig6_hFit}(a)--(d) displays the optimal fit results for \yi{} and \yc{}, and corresponding parameters are marked as orange stars in Fig. \ref{Fig6_hFit}(e)--(f): ($J_1$, $J_2$) = (0.043\,meV, 0.013\,meV) for \yi{} and (0.043\,meV, 0.010\,meV) for \yc{}. Regarding potential systematic errors included in this analysis, the fitted parameters contain uncertainty to some extent, which are roughly represented by the blue-colored regions in Fig. \ref{Fig6_hFit}(e)--(f). The range $ 0.26 \leq J_{2}/J_{1} \leq 0.37 $ ($ 0.16 \leq J_{2}/J_{1} \leq 0.34 $) expresses the uncertainty based on $2\chi_{\mathrm{min}}^{2}$ in \yi{} (\yc{}). However, even considering such uncertainty, our analysis clearly suggests sizable antiferromagnetic $J_{2}$ in both compounds, again indicating \yi{} and \yc{} are frustrated. In addition, we varied the fitted temperature ranges and only observed small variations of $J_{1}$ and $J_{2}$ values within the uncertainty illustrated as the blue-colored regions in Fig. \ref{Fig6_hFit}(e)--(f). 

Another noteworthy result is that the fitted $J_{2}/J_{1}$ of \yi{} ($\sim0.30$) is larger than that of \yc ($\sim0.23$), implying that \yi{} possesses stronger frustration. This can naturally describe the lower \TN{} and $T_{\mathrm{max}}$ of \yi{} than those of \yc{}, despite its larger (or nearly equal) $|$\CW{}$|$. However, we note that qualitatively the same tendency can be mimicked if \yc{} has stronger ferromagnetic $J_{c}$; this also increases \TN{} while decreasing $|$\CW{}$|$ to some extent. A shorter $c$ lattice parameter found in \yc{} (see Table \ref{tab_refine}) further supports this possibility. Yet the sign of $J_{c}$ cannot be determined by the results presented in this work. This requires a magnetic neutron diffraction measurement, which can confirm the sign of $J_{c}$ by revealing relative spin alignment between the Yb$^{3+}$ moments connected by $J_{c}$. Meanwhile, a comparison to exact-diagonalization (ED) results implies that our analysis might be underestimating $J_{2}/J_{1}$: according to ED, $|$\CW$|$/$T_{\mathrm{max}}=\,$1.92 found in \yc{} corresponds to $J_{2}/J_{1}$ larger than 0.25 \cite{Ref_Tmax2}. Therefore, the accurate values of $J_{2}/J_{1}$, as well as the validity of the $J_{1}$--$J_{2}$ model, should be confirmed by future measurements (\textit{e.g.}, inelastic neutron scattering). In particular, symmetry allowed anisotropic exchange interactions are another possible source of frustration, which are not considered in our isotropic spin model. Although recent studies suggest that Yb$^{3+}$-based quantum magnets often consist of Heisenberg type exchange interactions \cite{YbCl3_natc,TLAF_nearQSL}, future inelastic neutron scattering experiments are required to clarify this possibility.

Based on the parameters suggested by the high-temperature series expansion analysis, both compounds lie in the region of a classical Neel-type magnetic ground state. However, they are very close to the highly-frustrated regime in the $S=1/2$ $J_{1}$--$J_{2}$ square lattice model. This is evident in Fig. \ref{Fig6_hFit}(e)--(f); the fitted $J_{2}/J_{1}$ (especially for \yi{}) are not far from the suggested phase boundary line between the Neel order and spin liquid ($J_{2}=0.4J_{1}$). Thus, sizable quantum fluctuations are expected in these compounds. For instance, in a $S=1/2$ triangular lattice antiferromagnet, proximity to its spin-liquid region led to an observation of significant quantum entanglement between local moments and the dominance of fractionalized quasi-particles, even though a classical long-range order is present \cite{TLAF_nearQSL}. Similar exotic features might be anticipated in the highly-frustrated $S=1/2$ square lattice, thereby making \yi{} and \yc{} worthy of further investigation.

\section{Conclusion}
We have studied the magnetic properties of \yi{} and \yc{} by measuring their specific heat and magnetization. The results suggest that the Yb$^{3+}$ sublattice realizes an ideal $J_{\mathrm{eff}}=1/2$ square lattice by forming a well-isolated Kramers' doublet. The specific heat measurements reveal that \yi{} and \yc{} form long-range order below \TN{} = 0.21\,K and 0.25\,K, respectively. However, sizable spin correlations are evidenced at temperatures much higher than \TN, which is attributed to a combination of pronounced two dimensional character and frustration. \yi{} and \yc{} could be ideal platforms for investigating quantum effects of frustrated square lattice antiferromagnets, and thus this study motivates future work on the magnetic ground state and excitations in these and related phases.

\begin{acknowledgments}
 This research was supported by the U.S. Department of Energy, Office of Science, Basic Energy Sciences, Materials Science and Engineering Division. A portion of this research used resources at the Spallation Neutron Source, a DOE Office of Science User Facility operated by the Oak Ridge National Laboratory.
\end{acknowledgments}

%\bibliography{mainscript}% Produces the bibliography via BibTeX.

\begin{thebibliography}{41}%
\makeatletter
\providecommand \@ifxundefined [1]{%
 \@ifx{#1\undefined}
}%
\providecommand \@ifnum [1]{%
 \ifnum #1\expandafter \@firstoftwo
 \else \expandafter \@secondoftwo
 \fi
}%
\providecommand \@ifx [1]{%
 \ifx #1\expandafter \@firstoftwo
 \else \expandafter \@secondoftwo
 \fi
}%
\providecommand \natexlab [1]{#1}%
\providecommand \enquote  [1]{``#1''}%
\providecommand \bibnamefont  [1]{#1}%
\providecommand \bibfnamefont [1]{#1}%
\providecommand \citenamefont [1]{#1}%
\providecommand \href@noop [0]{\@secondoftwo}%
\providecommand \href [0]{\begingroup \@sanitize@url \@href}%
\providecommand \@href[1]{\@@startlink{#1}\@@href}%
\providecommand \@@href[1]{\endgroup#1\@@endlink}%
\providecommand \@sanitize@url [0]{\catcode `\\12\catcode `\$12\catcode
  `\&12\catcode `\#12\catcode `\^12\catcode `\_12\catcode `\%12\relax}%
\providecommand \@@startlink[1]{}%
\providecommand \@@endlink[0]{}%
\providecommand \url  [0]{\begingroup\@sanitize@url \@url }%
\providecommand \@url [1]{\endgroup\@href {#1}{\urlprefix }}%
\providecommand \urlprefix  [0]{URL }%
\providecommand \Eprint [0]{\href }%
\providecommand \doibase [0]{https://doi.org/}%
\providecommand \selectlanguage [0]{\@gobble}%
\providecommand \bibinfo  [0]{\@secondoftwo}%
\providecommand \bibfield  [0]{\@secondoftwo}%
\providecommand \translation [1]{[#1]}%
\providecommand \BibitemOpen [0]{}%
\providecommand \bibitemStop [0]{}%
\providecommand \bibitemNoStop [0]{.\EOS\space}%
\providecommand \EOS [0]{\spacefactor3000\relax}%
\providecommand \BibitemShut  [1]{\csname bibitem#1\endcsname}%
\let\auto@bib@innerbib\@empty
%</preamble>
\bibitem [{\citenamefont {Reger}\ and\ \citenamefont
  {Young}(1988)}]{Reger_1988}%
  \BibitemOpen
  \bibfield  {author} {\bibinfo {author} {\bibfnamefont {J.~D.}\ \bibnamefont
  {Reger}}\ and\ \bibinfo {author} {\bibfnamefont {A.~P.}\ \bibnamefont
  {Young}},\ }\bibfield  {title} {\bibinfo {title} {{Monte Carlo} simulations
  of the spin-1/2 {Heisenberg} antiferromagnet on a square lattice},\ }\href
  {https://doi.org/10.1103/PhysRevB.37.5978} {\bibfield  {journal} {\bibinfo
  {journal} {Physical Review B}\ }\textbf {\bibinfo {volume} {37}},\ \bibinfo
  {pages} {5978} (\bibinfo {year} {1988})}\BibitemShut {NoStop}%
\bibitem [{\citenamefont {Dagotto}\ and\ \citenamefont
  {Moreo}(1989)}]{Dagotto_1989}%
  \BibitemOpen
  \bibfield  {author} {\bibinfo {author} {\bibfnamefont {E.}~\bibnamefont
  {Dagotto}}\ and\ \bibinfo {author} {\bibfnamefont {A.}~\bibnamefont
  {Moreo}},\ }\bibfield  {title} {\bibinfo {title} {Phase diagram of the
  frustrated spin-1/2 heisenberg antiferromagnet in 2 dimensions},\ }\href
  {https://doi.org/10.1103/PhysRevLett.63.2148} {\bibfield  {journal} {\bibinfo
   {journal} {Phys. Rev. Lett.}\ }\textbf {\bibinfo {volume} {63}},\ \bibinfo
  {pages} {2148} (\bibinfo {year} {1989})}\BibitemShut {NoStop}%
\bibitem [{\citenamefont {Makivi{\'c}}\ and\ \citenamefont
  {Ding}(1991)}]{QMC_1991}%
  \BibitemOpen
  \bibfield  {author} {\bibinfo {author} {\bibfnamefont {M.~S.}\ \bibnamefont
  {Makivi{\'c}}}\ and\ \bibinfo {author} {\bibfnamefont {H.-Q.}\ \bibnamefont
  {Ding}},\ }\bibfield  {title} {\bibinfo {title} {Two-dimensional spin-1/2
  heisenberg antiferromagnet: A quantum monte carlo study},\ }\href@noop {}
  {\bibfield  {journal} {\bibinfo  {journal} {Physical Review B}\ }\textbf
  {\bibinfo {volume} {43}},\ \bibinfo {pages} {3562} (\bibinfo {year}
  {1991})}\BibitemShut {NoStop}%
\bibitem [{\citenamefont {Schulz}\ and\ \citenamefont
  {Ziman}(1992)}]{Schulz_1992}%
  \BibitemOpen
  \bibfield  {author} {\bibinfo {author} {\bibfnamefont {H.~J.}\ \bibnamefont
  {Schulz}}\ and\ \bibinfo {author} {\bibfnamefont {T.~A.~L.}\ \bibnamefont
  {Ziman}},\ }\bibfield  {title} {\bibinfo {title} {Finite-size scaling for the
  two-dimensional frustrated quantum heisenberg antiferromagnet},\ }\href
  {https://doi.org/10.1209/0295-5075/18/4/013} {\bibfield  {journal} {\bibinfo
  {journal} {Europhysics Letters}\ }\textbf {\bibinfo {volume} {18}},\ \bibinfo
  {pages} {355} (\bibinfo {year} {1992})}\BibitemShut {NoStop}%
\bibitem [{\citenamefont {Chandra}\ and\ \citenamefont
  {Doucot}(1988)}]{Chandra_1988}%
  \BibitemOpen
  \bibfield  {author} {\bibinfo {author} {\bibfnamefont {P.}~\bibnamefont
  {Chandra}}\ and\ \bibinfo {author} {\bibfnamefont {B.}~\bibnamefont
  {Doucot}},\ }\bibfield  {title} {\bibinfo {title} {Possible spin-liquid state
  at large {$S$} for the frustrated square {Heisenberg} lattice},\ }\href
  {https://doi.org/10.1103/PhysRevB.38.9335} {\bibfield  {journal} {\bibinfo
  {journal} {Phys. Rev. B}\ }\textbf {\bibinfo {volume} {38}},\ \bibinfo
  {pages} {9335} (\bibinfo {year} {1988})}\BibitemShut {NoStop}%
\bibitem [{\citenamefont {Manousakis}(1991)}]{RMP_1991}%
  \BibitemOpen
  \bibfield  {author} {\bibinfo {author} {\bibfnamefont {E.}~\bibnamefont
  {Manousakis}},\ }\bibfield  {title} {\bibinfo {title} {The spin-1/2
  heisenberg antiferromagnet on a square lattice and its application to the
  cuprous oxides},\ }\href {https://doi.org/10.1103/RevModPhys.63.1} {\bibfield
   {journal} {\bibinfo  {journal} {Reviews of Modern Physics}\ }\textbf
  {\bibinfo {volume} {63}},\ \bibinfo {pages} {1} (\bibinfo {year}
  {1991})}\BibitemShut {NoStop}%
\bibitem [{\citenamefont {Anderson}(1987)}]{Anderson_1987}%
  \BibitemOpen
  \bibfield  {author} {\bibinfo {author} {\bibfnamefont {P.~W.}\ \bibnamefont
  {Anderson}},\ }\bibfield  {title} {\bibinfo {title} {The resonating valence
  bond state in {La$_{2}$CuO$_{4}$} and superconductivity},\ }\href@noop {}
  {\bibfield  {journal} {\bibinfo  {journal} {science}\ }\textbf {\bibinfo
  {volume} {235}},\ \bibinfo {pages} {1196} (\bibinfo {year}
  {1987})}\BibitemShut {NoStop}%
\bibitem [{\citenamefont {Hamer}\ \emph {et~al.}(1992)\citenamefont {Hamer},
  \citenamefont {Zheng},\ and\ \citenamefont {Arndt}}]{Hamer_1992}%
  \BibitemOpen
  \bibfield  {author} {\bibinfo {author} {\bibfnamefont {C.~J.}\ \bibnamefont
  {Hamer}}, \bibinfo {author} {\bibfnamefont {W.}~\bibnamefont {Zheng}},\ and\
  \bibinfo {author} {\bibfnamefont {P.}~\bibnamefont {Arndt}},\ }\bibfield
  {title} {\bibinfo {title} {Third-order spin-wave theory for the {Heisenberg}
  antiferromagnet},\ }\href {https://doi.org/10.1103/PhysRevB.46.6276}
  {\bibfield  {journal} {\bibinfo  {journal} {Physical Review B}\ }\textbf
  {\bibinfo {volume} {46}},\ \bibinfo {pages} {6276} (\bibinfo {year}
  {1992})}\BibitemShut {NoStop}%
\bibitem [{\citenamefont {Dalla~Piazza}\ \emph {et~al.}(2015)\citenamefont
  {Dalla~Piazza}, \citenamefont {Mourigal}, \citenamefont {Christensen},
  \citenamefont {Nilsen}, \citenamefont {Tregenna-Piggott}, \citenamefont
  {Perring}, \citenamefont {Enderle}, \citenamefont {McMorrow}, \citenamefont
  {Ivanov},\ and\ \citenamefont {Rønnow}}]{Mourigal_2015}%
  \BibitemOpen
  \bibfield  {author} {\bibinfo {author} {\bibfnamefont {B.}~\bibnamefont
  {Dalla~Piazza}}, \bibinfo {author} {\bibfnamefont {M.}~\bibnamefont
  {Mourigal}}, \bibinfo {author} {\bibfnamefont {N.~B.}\ \bibnamefont
  {Christensen}}, \bibinfo {author} {\bibfnamefont {G.~J.}\ \bibnamefont
  {Nilsen}}, \bibinfo {author} {\bibfnamefont {P.}~\bibnamefont
  {Tregenna-Piggott}}, \bibinfo {author} {\bibfnamefont {T.~G.}\ \bibnamefont
  {Perring}}, \bibinfo {author} {\bibfnamefont {M.}~\bibnamefont {Enderle}},
  \bibinfo {author} {\bibfnamefont {D.~F.}\ \bibnamefont {McMorrow}}, \bibinfo
  {author} {\bibfnamefont {D.~A.}\ \bibnamefont {Ivanov}},\ and\ \bibinfo
  {author} {\bibfnamefont {H.~M.}\ \bibnamefont {Rønnow}},\ }\bibfield
  {title} {\bibinfo {title} {Fractional excitations in the square-lattice
  quantum antiferromagnet},\ }\href {https://doi.org/10.1038/nphys3172}
  {\bibfield  {journal} {\bibinfo  {journal} {Nature Physics}\ }\textbf
  {\bibinfo {volume} {11}},\ \bibinfo {pages} {62} (\bibinfo {year}
  {2015})}\BibitemShut {NoStop}%
\bibitem [{\citenamefont {Christensen}\ \emph {et~al.}(2007)\citenamefont
  {Christensen}, \citenamefont {Rønnow}, \citenamefont {McMorrow},
  \citenamefont {Harrison}, \citenamefont {Perring}, \citenamefont {Enderle},
  \citenamefont {Coldea}, \citenamefont {Regnault},\ and\ \citenamefont
  {Aeppli}}]{Christensen_2007}%
  \BibitemOpen
  \bibfield  {author} {\bibinfo {author} {\bibfnamefont {N.~B.}\ \bibnamefont
  {Christensen}}, \bibinfo {author} {\bibfnamefont {H.~M.}\ \bibnamefont
  {Rønnow}}, \bibinfo {author} {\bibfnamefont {D.~F.}\ \bibnamefont
  {McMorrow}}, \bibinfo {author} {\bibfnamefont {A.}~\bibnamefont {Harrison}},
  \bibinfo {author} {\bibfnamefont {T.~G.}\ \bibnamefont {Perring}}, \bibinfo
  {author} {\bibfnamefont {M.}~\bibnamefont {Enderle}}, \bibinfo {author}
  {\bibfnamefont {R.}~\bibnamefont {Coldea}}, \bibinfo {author} {\bibfnamefont
  {L.~P.}\ \bibnamefont {Regnault}},\ and\ \bibinfo {author} {\bibfnamefont
  {G.}~\bibnamefont {Aeppli}},\ }\bibfield  {title} {\bibinfo {title} {Quantum
  dynamics and entanglement of spins on a square lattice},\ }\href
  {https://doi.org/doi:10.1073/pnas.0703293104} {\bibfield  {journal} {\bibinfo
   {journal} {Proceedings of the National Academy of Sciences}\ }\textbf
  {\bibinfo {volume} {104}},\ \bibinfo {pages} {15264} (\bibinfo {year}
  {2007})}\BibitemShut {NoStop}%
\bibitem [{\citenamefont {Jiang}\ \emph {et~al.}(2012)\citenamefont {Jiang},
  \citenamefont {Yao},\ and\ \citenamefont {Balents}}]{Balents_2012}%
  \BibitemOpen
  \bibfield  {author} {\bibinfo {author} {\bibfnamefont {H.-C.}\ \bibnamefont
  {Jiang}}, \bibinfo {author} {\bibfnamefont {H.}~\bibnamefont {Yao}},\ and\
  \bibinfo {author} {\bibfnamefont {L.}~\bibnamefont {Balents}},\ }\bibfield
  {title} {\bibinfo {title} {Spin liquid ground state of the spin-$\frac{1}{2}$
  square {${J}_{1}$-${J}_{2}$} {Heisenberg} model},\ }\href
  {https://doi.org/10.1103/PhysRevB.86.024424} {\bibfield  {journal} {\bibinfo
  {journal} {Physical Review B}\ }\textbf {\bibinfo {volume} {86}},\ \bibinfo
  {pages} {024424} (\bibinfo {year} {2012})}\BibitemShut {NoStop}%
\bibitem [{\citenamefont {Morita}\ \emph {et~al.}(2015)\citenamefont {Morita},
  \citenamefont {Kaneko},\ and\ \citenamefont {Imada}}]{Morita_2015}%
  \BibitemOpen
  \bibfield  {author} {\bibinfo {author} {\bibfnamefont {S.}~\bibnamefont
  {Morita}}, \bibinfo {author} {\bibfnamefont {R.}~\bibnamefont {Kaneko}},\
  and\ \bibinfo {author} {\bibfnamefont {M.}~\bibnamefont {Imada}},\ }\bibfield
   {title} {\bibinfo {title} {Quantum spin liquid in spin 1/2 {$J_{1}–J_{2}$}
  {Heisenberg} model on square lattice: Many-variable variational {Monte Carlo}
  study combined with quantum-number projections},\ }\href
  {https://doi.org/10.7566/JPSJ.84.024720} {\bibfield  {journal} {\bibinfo
  {journal} {Journal of the Physical Society of Japan}\ }\textbf {\bibinfo
  {volume} {84}},\ \bibinfo {pages} {024720} (\bibinfo {year} {2015})},\
  \bibinfo {note} {doi: 10.7566/JPSJ.84.024720}\BibitemShut {NoStop}%
\bibitem [{\citenamefont {Gong}\ \emph {et~al.}(2014)\citenamefont {Gong},
  \citenamefont {Zhu}, \citenamefont {Sheng}, \citenamefont {Motrunich},\ and\
  \citenamefont {Fisher}}]{Gong_2014}%
  \BibitemOpen
  \bibfield  {author} {\bibinfo {author} {\bibfnamefont {S.-S.}\ \bibnamefont
  {Gong}}, \bibinfo {author} {\bibfnamefont {W.}~\bibnamefont {Zhu}}, \bibinfo
  {author} {\bibfnamefont {D.~N.}\ \bibnamefont {Sheng}}, \bibinfo {author}
  {\bibfnamefont {O.~I.}\ \bibnamefont {Motrunich}},\ and\ \bibinfo {author}
  {\bibfnamefont {M.~P.~A.}\ \bibnamefont {Fisher}},\ }\bibfield  {title}
  {\bibinfo {title} {Plaquette ordered phase and quantum phase diagram in the
  spin-{$\frac{1}{2}$ ${J}_{1}\text{\ensuremath{-}}{J}_{2}$} square
  {Heisenberg} model},\ }\href {https://doi.org/10.1103/PhysRevLett.113.027201}
  {\bibfield  {journal} {\bibinfo  {journal} {Physical Review Letters}\
  }\textbf {\bibinfo {volume} {113}},\ \bibinfo {pages} {027201} (\bibinfo
  {year} {2014})}\BibitemShut {NoStop}%
\bibitem [{\citenamefont {Liu}\ \emph {et~al.}(2022)\citenamefont {Liu},
  \citenamefont {Gong}, \citenamefont {Li}, \citenamefont {Poilblanc},
  \citenamefont {Chen},\ and\ \citenamefont {Gu}}]{Liu_2022_Tens}%
  \BibitemOpen
  \bibfield  {author} {\bibinfo {author} {\bibfnamefont {W.-Y.}\ \bibnamefont
  {Liu}}, \bibinfo {author} {\bibfnamefont {S.-S.}\ \bibnamefont {Gong}},
  \bibinfo {author} {\bibfnamefont {Y.-B.}\ \bibnamefont {Li}}, \bibinfo
  {author} {\bibfnamefont {D.}~\bibnamefont {Poilblanc}}, \bibinfo {author}
  {\bibfnamefont {W.-Q.}\ \bibnamefont {Chen}},\ and\ \bibinfo {author}
  {\bibfnamefont {Z.-C.}\ \bibnamefont {Gu}},\ }\bibfield  {title} {\bibinfo
  {title} {Gapless quantum spin liquid and global phase diagram of the spin-1/2
  {${\mathit{J}}_{1}-{\mathit{J}}_{2}$} square antiferromagnetic {Heisenberg}
  model},\ }\href {https://doi.org/https://doi.org/10.1016/j.scib.2022.03.010}
  {\bibfield  {journal} {\bibinfo  {journal} {Science Bulletin}\ }\textbf
  {\bibinfo {volume} {67}},\ \bibinfo {pages} {1034} (\bibinfo {year}
  {2022})}\BibitemShut {NoStop}%
\bibitem [{\citenamefont {Carretta}\ \emph {et~al.}(2002)\citenamefont
  {Carretta}, \citenamefont {Papinutto}, \citenamefont {Azzoni}, \citenamefont
  {Mozzati}, \citenamefont {Pavarini}, \citenamefont {Gonthier},\ and\
  \citenamefont {Millet}}]{Sq_ex_1}%
  \BibitemOpen
  \bibfield  {author} {\bibinfo {author} {\bibfnamefont {P.}~\bibnamefont
  {Carretta}}, \bibinfo {author} {\bibfnamefont {N.}~\bibnamefont {Papinutto}},
  \bibinfo {author} {\bibfnamefont {C.~B.}\ \bibnamefont {Azzoni}}, \bibinfo
  {author} {\bibfnamefont {M.~C.}\ \bibnamefont {Mozzati}}, \bibinfo {author}
  {\bibfnamefont {E.}~\bibnamefont {Pavarini}}, \bibinfo {author}
  {\bibfnamefont {S.}~\bibnamefont {Gonthier}},\ and\ \bibinfo {author}
  {\bibfnamefont {P.}~\bibnamefont {Millet}},\ }\bibfield  {title} {\bibinfo
  {title} {Frustration-driven structural distortion in
  {$\mathrm{VOMoO}_{4}$}},\ }\href {https://doi.org/10.1103/PhysRevB.66.094420}
  {\bibfield  {journal} {\bibinfo  {journal} {Physical Review B}\ }\textbf
  {\bibinfo {volume} {66}},\ \bibinfo {pages} {094420} (\bibinfo {year}
  {2002})}\BibitemShut {NoStop}%
\bibitem [{\citenamefont {Melzi}\ \emph {et~al.}(2000)\citenamefont {Melzi},
  \citenamefont {Carretta}, \citenamefont {Lascialfari}, \citenamefont
  {Mambrini}, \citenamefont {Troyer}, \citenamefont {Millet},\ and\
  \citenamefont {Mila}}]{Ref_Tmax2}%
  \BibitemOpen
  \bibfield  {author} {\bibinfo {author} {\bibfnamefont {R.}~\bibnamefont
  {Melzi}}, \bibinfo {author} {\bibfnamefont {P.}~\bibnamefont {Carretta}},
  \bibinfo {author} {\bibfnamefont {A.}~\bibnamefont {Lascialfari}}, \bibinfo
  {author} {\bibfnamefont {M.}~\bibnamefont {Mambrini}}, \bibinfo {author}
  {\bibfnamefont {M.}~\bibnamefont {Troyer}}, \bibinfo {author} {\bibfnamefont
  {P.}~\bibnamefont {Millet}},\ and\ \bibinfo {author} {\bibfnamefont
  {F.}~\bibnamefont {Mila}},\ }\bibfield  {title} {\bibinfo {title}
  {{${\mathrm{Li}}_{2}\mathrm{VO}\mathrm{(Si,Ge)O}_{4}$}, a prototype of a
  two-dimensional frustrated quantum heisenberg antiferromagnet},\ }\href
  {https://doi.org/10.1103/PhysRevLett.85.1318} {\bibfield  {journal} {\bibinfo
   {journal} {Phys. Rev. Lett.}\ }\textbf {\bibinfo {volume} {85}},\ \bibinfo
  {pages} {1318} (\bibinfo {year} {2000})}\BibitemShut {NoStop}%
\bibitem [{\citenamefont {Mustonen}\ \emph
  {et~al.}(2018{\natexlab{a}})\citenamefont {Mustonen}, \citenamefont {Vasala},
  \citenamefont {Sadrollahi}, \citenamefont {Schmidt}, \citenamefont {Baines},
  \citenamefont {Walker}, \citenamefont {Terasaki}, \citenamefont {Litterst},
  \citenamefont {Baggio-Saitovitch},\ and\ \citenamefont
  {Karppinen}}]{Sq_ex_3}%
  \BibitemOpen
  \bibfield  {author} {\bibinfo {author} {\bibfnamefont {O.}~\bibnamefont
  {Mustonen}}, \bibinfo {author} {\bibfnamefont {S.}~\bibnamefont {Vasala}},
  \bibinfo {author} {\bibfnamefont {E.}~\bibnamefont {Sadrollahi}}, \bibinfo
  {author} {\bibfnamefont {K.~P.}\ \bibnamefont {Schmidt}}, \bibinfo {author}
  {\bibfnamefont {C.}~\bibnamefont {Baines}}, \bibinfo {author} {\bibfnamefont
  {H.~C.}\ \bibnamefont {Walker}}, \bibinfo {author} {\bibfnamefont
  {I.}~\bibnamefont {Terasaki}}, \bibinfo {author} {\bibfnamefont {F.~J.}\
  \bibnamefont {Litterst}}, \bibinfo {author} {\bibfnamefont {E.}~\bibnamefont
  {Baggio-Saitovitch}},\ and\ \bibinfo {author} {\bibfnamefont
  {M.}~\bibnamefont {Karppinen}},\ }\bibfield  {title} {\bibinfo {title}
  {Spin-liquid-like state in a spin-1/2 square-lattice antiferromagnet
  perovskite induced by d10–d0 cation mixing},\ }\href
  {https://doi.org/10.1038/s41467-018-03435-1} {\bibfield  {journal} {\bibinfo
  {journal} {Nature Communications}\ }\textbf {\bibinfo {volume} {9}},\
  \bibinfo {pages} {1085} (\bibinfo {year} {2018}{\natexlab{a}})}\BibitemShut
  {NoStop}%
\bibitem [{\citenamefont {Vasala}\ \emph {et~al.}(2012)\citenamefont {Vasala},
  \citenamefont {Cheng}, \citenamefont {Yamauchi}, \citenamefont {Goodenough},\
  and\ \citenamefont {Karppinen}}]{Sq_ex_4}%
  \BibitemOpen
  \bibfield  {author} {\bibinfo {author} {\bibfnamefont {S.}~\bibnamefont
  {Vasala}}, \bibinfo {author} {\bibfnamefont {J.~G.}\ \bibnamefont {Cheng}},
  \bibinfo {author} {\bibfnamefont {H.}~\bibnamefont {Yamauchi}}, \bibinfo
  {author} {\bibfnamefont {J.~B.}\ \bibnamefont {Goodenough}},\ and\ \bibinfo
  {author} {\bibfnamefont {M.}~\bibnamefont {Karppinen}},\ }\bibfield  {title}
  {\bibinfo {title} {Synthesis and characterization of
  {$\mathrm{S}{\mathrm{r}}_{2}\mathrm{Cu}(\mathrm{W}_{1\text{\ensuremath{-}}x}{\mathrm{M}\mathrm{o}}_{x}){\mathrm{O}}_{6}$}:
  A quasi-two-dimensional magnetic system},\ }\href
  {https://doi.org/10.1021/cm301154n} {\bibfield  {journal} {\bibinfo
  {journal} {Chemistry of Materials}\ }\textbf {\bibinfo {volume} {24}},\
  \bibinfo {pages} {2764} (\bibinfo {year} {2012})},\ \bibinfo {note} {doi:
  10.1021/cm301154n}\BibitemShut {NoStop}%
\bibitem [{\citenamefont {Watanabe}\ \emph {et~al.}(2018)\citenamefont
  {Watanabe}, \citenamefont {Kurita}, \citenamefont {Tanaka}, \citenamefont
  {Ueno}, \citenamefont {Matsui},\ and\ \citenamefont {Goto}}]{Sq_ex_5}%
  \BibitemOpen
  \bibfield  {author} {\bibinfo {author} {\bibfnamefont {M.}~\bibnamefont
  {Watanabe}}, \bibinfo {author} {\bibfnamefont {N.}~\bibnamefont {Kurita}},
  \bibinfo {author} {\bibfnamefont {H.}~\bibnamefont {Tanaka}}, \bibinfo
  {author} {\bibfnamefont {W.}~\bibnamefont {Ueno}}, \bibinfo {author}
  {\bibfnamefont {K.}~\bibnamefont {Matsui}},\ and\ \bibinfo {author}
  {\bibfnamefont {T.}~\bibnamefont {Goto}},\ }\bibfield  {title} {\bibinfo
  {title} {Valence-bond-glass state with a singlet gap in the
  spin-$\frac{1}{2}$ square-lattice random
  {${J}_{1}\text{\ensuremath{-}}{J}_{2}$} heisenberg antiferromagnet
  {${\mathrm{Sr}}_{2}{\mathrm{CuTe}}_{1\ensuremath{-}x}{\mathrm{W}}_{x}{\mathrm{O}}_{6}$}},\
  }\href {https://doi.org/10.1103/PhysRevB.98.054422} {\bibfield  {journal}
  {\bibinfo  {journal} {Physical Review B}\ }\textbf {\bibinfo {volume} {98}},\
  \bibinfo {pages} {054422} (\bibinfo {year} {2018})}\BibitemShut {NoStop}%
\bibitem [{\citenamefont {McGuire}(2017)}]{TM_halide_2017}%
  \BibitemOpen
  \bibfield  {author} {\bibinfo {author} {\bibfnamefont {M.}~\bibnamefont
  {McGuire}},\ }\bibfield  {title} {\bibinfo {title} {Crystal and magnetic
  structures in layered, transition metal dihalides and trihalides},\ }\href
  {https://doi.org/10.3390/cryst7050121} {\bibfield  {journal} {\bibinfo
  {journal} {Crystals}\ }\textbf {\bibinfo {volume} {7}},\ \bibinfo {pages}
  {121} (\bibinfo {year} {2017})}\BibitemShut {NoStop}%
\bibitem [{\citenamefont {Huang}\ \emph {et~al.}(2017)\citenamefont {Huang},
  \citenamefont {Clark}, \citenamefont {Navarro-Moratalla}, \citenamefont
  {Klein}, \citenamefont {Cheng}, \citenamefont {Seyler}, \citenamefont
  {Zhong}, \citenamefont {Schmidgall}, \citenamefont {McGuire},\ and\
  \citenamefont {Cobden}}]{CrI3_nat}%
  \BibitemOpen
  \bibfield  {author} {\bibinfo {author} {\bibfnamefont {B.}~\bibnamefont
  {Huang}}, \bibinfo {author} {\bibfnamefont {G.}~\bibnamefont {Clark}},
  \bibinfo {author} {\bibfnamefont {E.}~\bibnamefont {Navarro-Moratalla}},
  \bibinfo {author} {\bibfnamefont {D.~R.}\ \bibnamefont {Klein}}, \bibinfo
  {author} {\bibfnamefont {R.}~\bibnamefont {Cheng}}, \bibinfo {author}
  {\bibfnamefont {K.~L.}\ \bibnamefont {Seyler}}, \bibinfo {author}
  {\bibfnamefont {D.}~\bibnamefont {Zhong}}, \bibinfo {author} {\bibfnamefont
  {E.}~\bibnamefont {Schmidgall}}, \bibinfo {author} {\bibfnamefont {M.~A.}\
  \bibnamefont {McGuire}},\ and\ \bibinfo {author} {\bibfnamefont {D.~H.}\
  \bibnamefont {Cobden}},\ }\bibfield  {title} {\bibinfo {title}
  {Layer-dependent ferromagnetism in a van der waals crystal down to the
  monolayer limit},\ }\href@noop {} {\bibfield  {journal} {\bibinfo  {journal}
  {Nature}\ }\textbf {\bibinfo {volume} {546}},\ \bibinfo {pages} {270}
  (\bibinfo {year} {2017})}\BibitemShut {NoStop}%
\bibitem [{\citenamefont {Song}\ \emph {et~al.}(2022)\citenamefont {Song},
  \citenamefont {Occhialini}, \citenamefont {Ergeçen}, \citenamefont {Ilyas},
  \citenamefont {Amoroso}, \citenamefont {Barone}, \citenamefont {Kapeghian},
  \citenamefont {Watanabe}, \citenamefont {Taniguchi},\ and\ \citenamefont
  {Botana}}]{NiI2_nat}%
  \BibitemOpen
  \bibfield  {author} {\bibinfo {author} {\bibfnamefont {Q.}~\bibnamefont
  {Song}}, \bibinfo {author} {\bibfnamefont {C.~A.}\ \bibnamefont
  {Occhialini}}, \bibinfo {author} {\bibfnamefont {E.}~\bibnamefont
  {Ergeçen}}, \bibinfo {author} {\bibfnamefont {B.}~\bibnamefont {Ilyas}},
  \bibinfo {author} {\bibfnamefont {D.}~\bibnamefont {Amoroso}}, \bibinfo
  {author} {\bibfnamefont {P.}~\bibnamefont {Barone}}, \bibinfo {author}
  {\bibfnamefont {J.}~\bibnamefont {Kapeghian}}, \bibinfo {author}
  {\bibfnamefont {K.}~\bibnamefont {Watanabe}}, \bibinfo {author}
  {\bibfnamefont {T.}~\bibnamefont {Taniguchi}},\ and\ \bibinfo {author}
  {\bibfnamefont {A.~S.}\ \bibnamefont {Botana}},\ }\bibfield  {title}
  {\bibinfo {title} {Evidence for a single-layer van der waals multiferroic},\
  }\href@noop {} {\bibfield  {journal} {\bibinfo  {journal} {Nature}\ }\textbf
  {\bibinfo {volume} {602}},\ \bibinfo {pages} {601} (\bibinfo {year}
  {2022})}\BibitemShut {NoStop}%
\bibitem [{\citenamefont {Ju}\ \emph {et~al.}(2021)\citenamefont {Ju},
  \citenamefont {Lee}, \citenamefont {Kim}, \citenamefont {Choi}, \citenamefont
  {Roh}, \citenamefont {Son}, \citenamefont {Park}, \citenamefont {Kim},
  \citenamefont {Jung},\ and\ \citenamefont {Kim}}]{NiI2_bilayer}%
  \BibitemOpen
  \bibfield  {author} {\bibinfo {author} {\bibfnamefont {H.}~\bibnamefont
  {Ju}}, \bibinfo {author} {\bibfnamefont {Y.}~\bibnamefont {Lee}}, \bibinfo
  {author} {\bibfnamefont {K.-T.}\ \bibnamefont {Kim}}, \bibinfo {author}
  {\bibfnamefont {I.~H.}\ \bibnamefont {Choi}}, \bibinfo {author}
  {\bibfnamefont {C.~J.}\ \bibnamefont {Roh}}, \bibinfo {author} {\bibfnamefont
  {S.}~\bibnamefont {Son}}, \bibinfo {author} {\bibfnamefont {P.}~\bibnamefont
  {Park}}, \bibinfo {author} {\bibfnamefont {J.~H.}\ \bibnamefont {Kim}},
  \bibinfo {author} {\bibfnamefont {T.~S.}\ \bibnamefont {Jung}},\ and\
  \bibinfo {author} {\bibfnamefont {J.~H.}\ \bibnamefont {Kim}},\ }\bibfield
  {title} {\bibinfo {title} {Possible persistence of multiferroic order down to
  bilayer limit of van der waals material {$\mathrm{NiI}_{2}$}},\ }\href@noop
  {} {\bibfield  {journal} {\bibinfo  {journal} {Nano letters}\ }\textbf
  {\bibinfo {volume} {21}},\ \bibinfo {pages} {5126} (\bibinfo {year}
  {2021})}\BibitemShut {NoStop}%
\bibitem [{\citenamefont {Burch}\ \emph {et~al.}(2018)\citenamefont {Burch},
  \citenamefont {Mandrus},\ and\ \citenamefont {Park}}]{burch2018magnetism}%
  \BibitemOpen
  \bibfield  {author} {\bibinfo {author} {\bibfnamefont {K.~S.}\ \bibnamefont
  {Burch}}, \bibinfo {author} {\bibfnamefont {D.}~\bibnamefont {Mandrus}},\
  and\ \bibinfo {author} {\bibfnamefont {J.-G.}\ \bibnamefont {Park}},\
  }\bibfield  {title} {\bibinfo {title} {Magnetism in two-dimensional van der
  waals materials},\ }\href@noop {} {\bibfield  {journal} {\bibinfo  {journal}
  {Nature}\ }\textbf {\bibinfo {volume} {563}},\ \bibinfo {pages} {47}
  (\bibinfo {year} {2018})}\BibitemShut {NoStop}%
\bibitem [{\citenamefont {Huang}\ \emph {et~al.}(2020)\citenamefont {Huang},
  \citenamefont {McGuire}, \citenamefont {May}, \citenamefont {Xiao},
  \citenamefont {Jarillo-Herrero},\ and\ \citenamefont
  {Xu}}]{huang2020emergent}%
  \BibitemOpen
  \bibfield  {author} {\bibinfo {author} {\bibfnamefont {B.}~\bibnamefont
  {Huang}}, \bibinfo {author} {\bibfnamefont {M.~A.}\ \bibnamefont {McGuire}},
  \bibinfo {author} {\bibfnamefont {A.~F.}\ \bibnamefont {May}}, \bibinfo
  {author} {\bibfnamefont {D.}~\bibnamefont {Xiao}}, \bibinfo {author}
  {\bibfnamefont {P.}~\bibnamefont {Jarillo-Herrero}},\ and\ \bibinfo {author}
  {\bibfnamefont {X.}~\bibnamefont {Xu}},\ }\bibfield  {title} {\bibinfo
  {title} {Emergent phenomena and proximity effects in two-dimensional magnets
  and heterostructures},\ }\href@noop {} {\bibfield  {journal} {\bibinfo
  {journal} {Nature Materials}\ }\textbf {\bibinfo {volume} {19}},\ \bibinfo
  {pages} {1276} (\bibinfo {year} {2020})}\BibitemShut {NoStop}%
\bibitem [{\citenamefont {Banerjee}\ \emph {et~al.}(2017)\citenamefont
  {Banerjee}, \citenamefont {Yan}, \citenamefont {Knolle}, \citenamefont
  {Bridges}, \citenamefont {Stone}, \citenamefont {Lumsden}, \citenamefont
  {Mandrus}, \citenamefont {Tennant}, \citenamefont {Moessner},\ and\
  \citenamefont {Nagler}}]{RuCl3_sci}%
  \BibitemOpen
  \bibfield  {author} {\bibinfo {author} {\bibfnamefont {A.}~\bibnamefont
  {Banerjee}}, \bibinfo {author} {\bibfnamefont {J.}~\bibnamefont {Yan}},
  \bibinfo {author} {\bibfnamefont {J.}~\bibnamefont {Knolle}}, \bibinfo
  {author} {\bibfnamefont {C.~A.}\ \bibnamefont {Bridges}}, \bibinfo {author}
  {\bibfnamefont {M.~B.}\ \bibnamefont {Stone}}, \bibinfo {author}
  {\bibfnamefont {M.~D.}\ \bibnamefont {Lumsden}}, \bibinfo {author}
  {\bibfnamefont {D.~G.}\ \bibnamefont {Mandrus}}, \bibinfo {author}
  {\bibfnamefont {D.~A.}\ \bibnamefont {Tennant}}, \bibinfo {author}
  {\bibfnamefont {R.}~\bibnamefont {Moessner}},\ and\ \bibinfo {author}
  {\bibfnamefont {S.~E.}\ \bibnamefont {Nagler}},\ }\bibfield  {title}
  {\bibinfo {title} {Neutron scattering in the proximate quantum spin liquid
  {$\alpha-\mathrm{RuCl}_{3}$}},\ }\href@noop {} {\bibfield  {journal}
  {\bibinfo  {journal} {Science}\ }\textbf {\bibinfo {volume} {356}},\ \bibinfo
  {pages} {1055} (\bibinfo {year} {2017})}\BibitemShut {NoStop}%
\bibitem [{\citenamefont {Do}\ \emph {et~al.}(2017)\citenamefont {Do},
  \citenamefont {Park}, \citenamefont {Yoshitake}, \citenamefont {Nasu},
  \citenamefont {Motome}, \citenamefont {Kwon}, \citenamefont {Adroja},
  \citenamefont {Voneshen}, \citenamefont {Kim},\ and\ \citenamefont
  {Jang}}]{RuCl3_nphys}%
  \BibitemOpen
  \bibfield  {author} {\bibinfo {author} {\bibfnamefont {S.-H.}\ \bibnamefont
  {Do}}, \bibinfo {author} {\bibfnamefont {S.-Y.}\ \bibnamefont {Park}},
  \bibinfo {author} {\bibfnamefont {J.}~\bibnamefont {Yoshitake}}, \bibinfo
  {author} {\bibfnamefont {J.}~\bibnamefont {Nasu}}, \bibinfo {author}
  {\bibfnamefont {Y.}~\bibnamefont {Motome}}, \bibinfo {author} {\bibfnamefont
  {Y.~S.}\ \bibnamefont {Kwon}}, \bibinfo {author} {\bibfnamefont
  {D.}~\bibnamefont {Adroja}}, \bibinfo {author} {\bibfnamefont
  {D.}~\bibnamefont {Voneshen}}, \bibinfo {author} {\bibfnamefont
  {K.}~\bibnamefont {Kim}},\ and\ \bibinfo {author} {\bibfnamefont {T.-H.}\
  \bibnamefont {Jang}},\ }\bibfield  {title} {\bibinfo {title} {Majorana
  fermions in the kitaev quantum spin system {$\alpha-\mathrm{RuCl}_{3}$}},\
  }\href@noop {} {\bibfield  {journal} {\bibinfo  {journal} {Nature Physics}\
  }\textbf {\bibinfo {volume} {13}},\ \bibinfo {pages} {1079} (\bibinfo {year}
  {2017})}\BibitemShut {NoStop}%
\bibitem [{\citenamefont {Sala}\ \emph {et~al.}(2021)\citenamefont {Sala},
  \citenamefont {Stone}, \citenamefont {Rai}, \citenamefont {May},
  \citenamefont {Laurell}, \citenamefont {Garlea}, \citenamefont {Butch},
  \citenamefont {Lumsden}, \citenamefont {Ehlers},\ and\ \citenamefont
  {Pokharel}}]{YbCl3_natc}%
  \BibitemOpen
  \bibfield  {author} {\bibinfo {author} {\bibfnamefont {G.}~\bibnamefont
  {Sala}}, \bibinfo {author} {\bibfnamefont {M.~B.}\ \bibnamefont {Stone}},
  \bibinfo {author} {\bibfnamefont {B.~K.}\ \bibnamefont {Rai}}, \bibinfo
  {author} {\bibfnamefont {A.~F.}\ \bibnamefont {May}}, \bibinfo {author}
  {\bibfnamefont {P.}~\bibnamefont {Laurell}}, \bibinfo {author} {\bibfnamefont
  {V.~O.}\ \bibnamefont {Garlea}}, \bibinfo {author} {\bibfnamefont {N.~P.}\
  \bibnamefont {Butch}}, \bibinfo {author} {\bibfnamefont {M.~D.}\ \bibnamefont
  {Lumsden}}, \bibinfo {author} {\bibfnamefont {G.}~\bibnamefont {Ehlers}},\
  and\ \bibinfo {author} {\bibfnamefont {G.}~\bibnamefont {Pokharel}},\
  }\bibfield  {title} {\bibinfo {title} {Van hove singularity in the magnon
  spectrum of the antiferromagnetic quantum honeycomb lattice},\ }\href@noop {}
  {\bibfield  {journal} {\bibinfo  {journal} {Nature communications}\ }\textbf
  {\bibinfo {volume} {12}},\ \bibinfo {pages} {171} (\bibinfo {year}
  {2021})}\BibitemShut {NoStop}%
\bibitem [{\citenamefont {Kim}\ \emph {et~al.}(2023)\citenamefont {Kim},
  \citenamefont {Kim}, \citenamefont {Park}, \citenamefont {Kim}, \citenamefont
  {Jeong}, \citenamefont {Ohira-Kawamura}, \citenamefont {Murai}, \citenamefont
  {Nakajima}, \citenamefont {Chernyshev}, \citenamefont {Mourigal} \emph
  {et~al.}}]{CoI2_nphys}%
  \BibitemOpen
  \bibfield  {author} {\bibinfo {author} {\bibfnamefont {C.}~\bibnamefont
  {Kim}}, \bibinfo {author} {\bibfnamefont {S.}~\bibnamefont {Kim}}, \bibinfo
  {author} {\bibfnamefont {P.}~\bibnamefont {Park}}, \bibinfo {author}
  {\bibfnamefont {T.}~\bibnamefont {Kim}}, \bibinfo {author} {\bibfnamefont
  {J.}~\bibnamefont {Jeong}}, \bibinfo {author} {\bibfnamefont
  {S.}~\bibnamefont {Ohira-Kawamura}}, \bibinfo {author} {\bibfnamefont
  {N.}~\bibnamefont {Murai}}, \bibinfo {author} {\bibfnamefont
  {K.}~\bibnamefont {Nakajima}}, \bibinfo {author} {\bibfnamefont
  {A.}~\bibnamefont {Chernyshev}}, \bibinfo {author} {\bibfnamefont
  {M.}~\bibnamefont {Mourigal}}, \emph {et~al.},\ }\bibfield  {title} {\bibinfo
  {title} {Bond-dependent anisotropy and magnon decay in cobalt-based kitaev
  triangular antiferromagnet},\ }\href@noop {} {\bibfield  {journal} {\bibinfo
  {journal} {Nature Physics}\ }\textbf {\bibinfo {volume} {19}},\ \bibinfo
  {pages} {1624} (\bibinfo {year} {2023})}\BibitemShut {NoStop}%
\bibitem [{\citenamefont {Inosov}(2018)}]{qmag_review}%
  \BibitemOpen
  \bibfield  {author} {\bibinfo {author} {\bibfnamefont {D.}~\bibnamefont
  {Inosov}},\ }\bibfield  {title} {\bibinfo {title} {Quantum magnetism in
  minerals},\ }\href@noop {} {\bibfield  {journal} {\bibinfo  {journal}
  {Advances in Physics}\ }\textbf {\bibinfo {volume} {67}},\ \bibinfo {pages}
  {149} (\bibinfo {year} {2018})}\BibitemShut {NoStop}%
\bibitem [{\citenamefont {Momma}\ and\ \citenamefont {Izumi}(2011)}]{vesta}%
  \BibitemOpen
  \bibfield  {author} {\bibinfo {author} {\bibfnamefont {K.}~\bibnamefont
  {Momma}}\ and\ \bibinfo {author} {\bibfnamefont {F.}~\bibnamefont {Izumi}},\
  }\bibfield  {title} {\bibinfo {title} {{{\it VESTA3} for three-dimensional
  visualization of crystal, volumetric and morphology data}},\ }\href
  {https://doi.org/10.1107/S0021889811038970} {\bibfield  {journal} {\bibinfo
  {journal} {Journal of Applied Crystallography}\ }\textbf {\bibinfo {volume}
  {44}},\ \bibinfo {pages} {1272} (\bibinfo {year} {2011})}\BibitemShut
  {NoStop}%
\bibitem [{\citenamefont {Schmidt}\ \emph {et~al.}(2000)\citenamefont
  {Schmidt}, \citenamefont {Oppermann}, \citenamefont {Hennig}, \citenamefont
  {Henn}, \citenamefont {Gmelin}, \citenamefont {Söger},\ and\ \citenamefont
  {Binnewies}}]{Schmidt_2000_structure}%
  \BibitemOpen
  \bibfield  {author} {\bibinfo {author} {\bibfnamefont {M.}~\bibnamefont
  {Schmidt}}, \bibinfo {author} {\bibfnamefont {H.}~\bibnamefont {Oppermann}},
  \bibinfo {author} {\bibfnamefont {C.}~\bibnamefont {Hennig}}, \bibinfo
  {author} {\bibfnamefont {R.}~\bibnamefont {Henn}}, \bibinfo {author}
  {\bibfnamefont {E.}~\bibnamefont {Gmelin}}, \bibinfo {author} {\bibfnamefont
  {N.}~\bibnamefont {Söger}},\ and\ \bibinfo {author} {\bibfnamefont
  {M.}~\bibnamefont {Binnewies}},\ }\bibfield  {title} {\bibinfo {title}
  {Investigations on the bismuth rare-earth oxyhalides
  {$\mathrm{Bi}_{2}\mathrm{REO}_{4}\mathrm{X}$} ({$\mathrm{X = Cl, Br, I}$})},\
  }\href@noop {} {\bibfield  {journal} {\bibinfo  {journal} {Zeitschrift fur
  Anorganische und Allgemeine Chemie}\ }\textbf {\bibinfo {volume} {626}},\
  \bibinfo {pages} {125} (\bibinfo {year} {2000})}\BibitemShut {NoStop}%
\bibitem [{\citenamefont {Huq}\ \emph {et~al.}(2011)\citenamefont {Huq},
  \citenamefont {Hodges}, \citenamefont {Heroux},\ and\ \citenamefont
  {Gourdon}}]{powgen2011}%
  \BibitemOpen
  \bibfield  {author} {\bibinfo {author} {\bibfnamefont {A.}~\bibnamefont
  {Huq}}, \bibinfo {author} {\bibfnamefont {J.~P.}\ \bibnamefont {Hodges}},
  \bibinfo {author} {\bibfnamefont {L.}~\bibnamefont {Heroux}},\ and\ \bibinfo
  {author} {\bibfnamefont {O.}~\bibnamefont {Gourdon}},\ }\bibfield  {title}
  {\bibinfo {title} {Powgen: a third-generation high resolution high-throughput
  powder diffraction instrument at the spallation neutron source},\ }\href@noop
  {} {\bibfield  {journal} {\bibinfo  {journal} {Zeitschrift fur
  Kristallographie Proceedings}\ }\textbf {\bibinfo {volume} {1}},\ \bibinfo
  {pages} {127} (\bibinfo {year} {2011})}\BibitemShut {NoStop}%
\bibitem [{\citenamefont {Rodriguez-Carvajal}(1993)}]{fullprof}%
  \BibitemOpen
  \bibfield  {author} {\bibinfo {author} {\bibfnamefont {J.}~\bibnamefont
  {Rodriguez-Carvajal}},\ }\bibfield  {title} {\bibinfo {title} {Recent
  advances in magnetic structure determination by neutron powder diffraction},\
  }\href@noop {} {\bibfield  {journal} {\bibinfo  {journal} {Physica B}\
  }\textbf {\bibinfo {volume} {192}},\ \bibinfo {pages} {55} (\bibinfo {year}
  {1993})}\BibitemShut {NoStop}%
\bibitem [{\citenamefont {Mustonen}\ \emph
  {et~al.}(2018{\natexlab{b}})\citenamefont {Mustonen}, \citenamefont {Vasala},
  \citenamefont {Schmidt}, \citenamefont {Sadrollahi}, \citenamefont {Walker},
  \citenamefont {Terasaki}, \citenamefont {Litterst}, \citenamefont
  {Baggio-Saitovitch},\ and\ \citenamefont {Karppinen}}]{Mustonen_2018}%
  \BibitemOpen
  \bibfield  {author} {\bibinfo {author} {\bibfnamefont {O.}~\bibnamefont
  {Mustonen}}, \bibinfo {author} {\bibfnamefont {S.}~\bibnamefont {Vasala}},
  \bibinfo {author} {\bibfnamefont {K.~P.}\ \bibnamefont {Schmidt}}, \bibinfo
  {author} {\bibfnamefont {E.}~\bibnamefont {Sadrollahi}}, \bibinfo {author}
  {\bibfnamefont {H.~C.}\ \bibnamefont {Walker}}, \bibinfo {author}
  {\bibfnamefont {I.}~\bibnamefont {Terasaki}}, \bibinfo {author}
  {\bibfnamefont {F.~J.}\ \bibnamefont {Litterst}}, \bibinfo {author}
  {\bibfnamefont {E.}~\bibnamefont {Baggio-Saitovitch}},\ and\ \bibinfo
  {author} {\bibfnamefont {M.}~\bibnamefont {Karppinen}},\ }\bibfield  {title}
  {\bibinfo {title} {Tuning the {$S=1/2$} square-lattice antiferromagnet
  {$\mathrm{S}{\mathrm{r}}_{2}\mathrm{Cu}(\mathrm{T}{\mathrm{e}}_{1\text{\ensuremath{-}}x}{\mathrm{W}}_{x}){\mathrm{O}}_{6}$}
  from n\'eel order to quantum disorder to columnar order},\ }\href
  {https://doi.org/10.1103/PhysRevB.98.064411} {\bibfield  {journal} {\bibinfo
  {journal} {Physical Review B}\ }\textbf {\bibinfo {volume} {98}},\ \bibinfo
  {pages} {064411} (\bibinfo {year} {2018}{\natexlab{b}})}\BibitemShut
  {NoStop}%
\bibitem [{\citenamefont {Kini}\ \emph {et~al.}(2006)\citenamefont {Kini},
  \citenamefont {Kaul},\ and\ \citenamefont {Geibel}}]{ZVPO}%
  \BibitemOpen
  \bibfield  {author} {\bibinfo {author} {\bibfnamefont {N.}~\bibnamefont
  {Kini}}, \bibinfo {author} {\bibfnamefont {E.}~\bibnamefont {Kaul}},\ and\
  \bibinfo {author} {\bibfnamefont {C.}~\bibnamefont {Geibel}},\ }\bibfield
  {title} {\bibinfo {title} {{$\mathrm{Zn_{2}VO(PO_{4})_{2}}$}: an {$S$}= 1/2
  heisenberg antiferromagnetic square lattice system},\ }\href@noop {}
  {\bibfield  {journal} {\bibinfo  {journal} {Journal of Physics: Condensed
  Matter}\ }\textbf {\bibinfo {volume} {18}},\ \bibinfo {pages} {1303}
  (\bibinfo {year} {2006})}\BibitemShut {NoStop}%
\bibitem [{\citenamefont {Bałanda}(2013)}]{Ref_AC}%
  \BibitemOpen
  \bibfield  {author} {\bibinfo {author} {\bibfnamefont {M.}~\bibnamefont
  {Bałanda}},\ }\bibfield  {title} {\bibinfo {title} {Ac susceptibility
  studies of phase transitions and magnetic relaxation: Conventional, molecular
  and low-dimensional magnets},\ }\href
  {https://doi.org/10.12693/APhysPolA.124.964} {\bibfield  {journal} {\bibinfo
  {journal} {Acta Physica Polonica A}\ }\textbf {\bibinfo {volume} {124}},\
  \bibinfo {pages} {964} (\bibinfo {year} {2013})}\BibitemShut {NoStop}%
\bibitem [{\citenamefont {Kageyama}\ \emph {et~al.}(2005)\citenamefont
  {Kageyama}, \citenamefont {Kitano}, \citenamefont {Oba}, \citenamefont
  {Nishi}, \citenamefont {Nagai}, \citenamefont {Hirota}, \citenamefont
  {Viciu}, \citenamefont {JB}, \citenamefont {Yasuda}, \citenamefont {Baba}
  \emph {et~al.}}]{Ref_Tmax1}%
  \BibitemOpen
  \bibfield  {author} {\bibinfo {author} {\bibfnamefont {H.}~\bibnamefont
  {Kageyama}}, \bibinfo {author} {\bibfnamefont {T.}~\bibnamefont {Kitano}},
  \bibinfo {author} {\bibfnamefont {N.}~\bibnamefont {Oba}}, \bibinfo {author}
  {\bibfnamefont {M.}~\bibnamefont {Nishi}}, \bibinfo {author} {\bibfnamefont
  {S.}~\bibnamefont {Nagai}}, \bibinfo {author} {\bibfnamefont
  {K.}~\bibnamefont {Hirota}}, \bibinfo {author} {\bibfnamefont
  {L.}~\bibnamefont {Viciu}}, \bibinfo {author} {\bibfnamefont
  {W.}~\bibnamefont {JB}}, \bibinfo {author} {\bibfnamefont {J.}~\bibnamefont
  {Yasuda}}, \bibinfo {author} {\bibfnamefont {Y.}~\bibnamefont {Baba}}, \emph
  {et~al.},\ }\bibfield  {title} {\bibinfo {title} {Spin-singlet ground state
  in two-dimensional {$S=1/2$} frustrated square
  lattice:{$\mathrm{(CuCl)LaNb}_{2}\mathrm{O}_{7}$}},\ }\href@noop {}
  {\bibfield  {journal} {\bibinfo  {journal} {Journal of the Physical Society
  of Japan}\ }\textbf {\bibinfo {volume} {74}},\ \bibinfo {pages} {1702}
  (\bibinfo {year} {2005})}\BibitemShut {NoStop}%
\bibitem [{\citenamefont {Schmidt}\ \emph {et~al.}(2011)\citenamefont
  {Schmidt}, \citenamefont {Lohmann},\ and\ \citenamefont
  {Richter}}]{highT_exp}%
  \BibitemOpen
  \bibfield  {author} {\bibinfo {author} {\bibfnamefont {H.-J.}\ \bibnamefont
  {Schmidt}}, \bibinfo {author} {\bibfnamefont {A.}~\bibnamefont {Lohmann}},\
  and\ \bibinfo {author} {\bibfnamefont {J.}~\bibnamefont {Richter}},\
  }\bibfield  {title} {\bibinfo {title} {Eighth-order high-temperature
  expansion for general heisenberg hamiltonians},\ }\href
  {https://doi.org/10.1103/PhysRevB.84.104443} {\bibfield  {journal} {\bibinfo
  {journal} {Phys. Rev. B}\ }\textbf {\bibinfo {volume} {84}},\ \bibinfo
  {pages} {104443} (\bibinfo {year} {2011})}\BibitemShut {NoStop}%
\bibitem [{\citenamefont {Moon}\ \emph {et~al.}(1967)\citenamefont {Moon},
  \citenamefont {Child}, \citenamefont {Koehler},\ and\ \citenamefont
  {Raubenheimer}}]{Yb2O3}%
  \BibitemOpen
  \bibfield  {author} {\bibinfo {author} {\bibfnamefont {R.}~\bibnamefont
  {Moon}}, \bibinfo {author} {\bibfnamefont {H.}~\bibnamefont {Child}},
  \bibinfo {author} {\bibfnamefont {W.}~\bibnamefont {Koehler}},\ and\ \bibinfo
  {author} {\bibfnamefont {L.}~\bibnamefont {Raubenheimer}},\ }\bibfield
  {title} {\bibinfo {title} {Magnetic structure of
  {$\mathrm{Er}_{2}\mathrm{O}_{3}$} and {$\mathrm{Yb}_{2}\mathrm{O}_{3}$}},\
  }\href@noop {} {\bibfield  {journal} {\bibinfo  {journal} {Journal of Applied
  Physics}\ }\textbf {\bibinfo {volume} {38}},\ \bibinfo {pages} {1383}
  (\bibinfo {year} {1967})}\BibitemShut {NoStop}%
\bibitem [{\citenamefont {Scheie}\ \emph {et~al.}(2023)\citenamefont {Scheie},
  \citenamefont {Ghioldi}, \citenamefont {Xing}, \citenamefont {Paddison},
  \citenamefont {Sherman}, \citenamefont {Dupont}, \citenamefont {Sanjeewa},
  \citenamefont {Lee}, \citenamefont {Woods}, \citenamefont {Abernathy} \emph
  {et~al.}}]{TLAF_nearQSL}%
  \BibitemOpen
  \bibfield  {author} {\bibinfo {author} {\bibfnamefont {A.}~\bibnamefont
  {Scheie}}, \bibinfo {author} {\bibfnamefont {E.}~\bibnamefont {Ghioldi}},
  \bibinfo {author} {\bibfnamefont {J.}~\bibnamefont {Xing}}, \bibinfo {author}
  {\bibfnamefont {J.}~\bibnamefont {Paddison}}, \bibinfo {author}
  {\bibfnamefont {N.}~\bibnamefont {Sherman}}, \bibinfo {author} {\bibfnamefont
  {M.}~\bibnamefont {Dupont}}, \bibinfo {author} {\bibfnamefont
  {L.}~\bibnamefont {Sanjeewa}}, \bibinfo {author} {\bibfnamefont
  {S.}~\bibnamefont {Lee}}, \bibinfo {author} {\bibfnamefont {A.}~\bibnamefont
  {Woods}}, \bibinfo {author} {\bibfnamefont {D.}~\bibnamefont {Abernathy}},
  \emph {et~al.},\ }\bibfield  {title} {\bibinfo {title} {Proximate spin liquid
  and fractionalization in the triangular antiferromagnet
  {$\mathrm{KYbSe_{2}}$}},\ }\href@noop {} {\bibfield  {journal} {\bibinfo
  {journal} {Nature Physics}\ ,\ \bibinfo {pages}
  {https://doi.org/10.1038/s41567}} (\bibinfo {year} {2023})}\BibitemShut
  {NoStop}%
\end{thebibliography}

\providecommand{\noopsort}[1]{}\providecommand{\singleletter}[1]{#1}%
%

% \appendix

\clearpage
\onecolumngrid

\renewcommand{\thefigure}{A\arabic{figure}}
\renewcommand{\thetable}{A\Roman{table}}
\setcounter{figure}{0}
\setcounter{table}{0}

\end{document}